\documentstyle[12pt,epsfig]{article}
\renewcommand{\baselinestretch}{2.0}
\begin{document}
\newcommand {\be}{\begin{equation}}
\newcommand {\ee}{\end{equation}}
\newcommand {\ba}{\begin{eqnarray}}
\newcommand {\ea}{\end{eqnarray}}
\newcommand {\bea}{\begin{array}}
\newcommand {\cl}{\centerline}

\newcommand {\eea}{\end{array}}
\renewcommand {\thefootnote}{\fnsymbol{footnote}}

\vskip .5cm

\renewcommand {\thefootnote}{\fnsymbol{footnote}}
\def \a'{\alpha'}
\baselineskip 0.65 cm
\baselineskip 0.65 cm
\begin{flushright}
SISSA/ 69/2002/EP\\
SLAC-PUB-9543 \\
hep-ph/0211375 \\
\today
\end{flushright}
\begin{center}
{\Large{\bf Bounds on the Coupling of the Majoron to Light Neutrinos 
 from 
Supernova Cooling \footnote{Work supported, in part, by the U. S. 
Department 
of Energy under
contract DE-AC03-76SF00515.} }}
{\vskip 0.5 cm}

{\bf {Yasaman  Farzan\\}}
{\vskip 0.5 cm}
{ 
{\it Stanford Linear Accelerator Center, 
                                              2575 Sand Hill Road,
                                             Menlo Park, CA 94025
 }}
{\vskip 0.5 cm}
and \\
{  {\it Scuola 
Internazionale superiore di Studi Avanzati,  via Beirut 4, I-34014 
Trieste, Italy}}
\end{center}
\begin{abstract}
We explore the role of Majoron ($J$) emission in the supernova cooling 
process, as a source of the upper bound on  neutrino-Majoron coupling.
 We show that the strongest upper bound on the coupling to $\nu_e$
 comes from the $\nu_e\nu_e \to J $ process in the core of 
a supernova. We also find bounds on diagonal couplings of the Majoron to 
$\nu_{\mu(\tau)}\nu_{\mu(\tau)}$ and on off-diagonal $\nu_e \nu_{\mu 
(\tau)}$
 couplings in  various regions of the parameter space.  
We discuss the evaluation of the cross-section  for four-particle 
interactions ($\nu  \nu \to JJ$ and  $\nu J \to \nu J$). We show 
that these are typically dominated by three-particle sub-processes and do 
not give new independent constraints. 
  \end{abstract}
 \section{Introduction}
The solar and atmospheric neutrino observations provide 
 strong evidence in 
favor of neutrinos being massive. These experiments are  sensitive 
only to 
mass squared differences \cite{sun}, 
on the other hand,
the overall mass scale of neutrinos is strongly constrained  
by the Troitsk and Mainz experiments \cite{mainz}. Combining 
these pieces of information, we conclude that the masses of the three 
active neutrinos are very small.
Among the  plausible and economic models which are developed
to give a tiny mass to neutrinos are Majoron models \cite{peccei,gr}. In 
these models, additional Higgs boson(s)  are introduced such 
that their 
vacuum expectation values break the exact B$-$L symmetry of the model. The 
Goldstone boson 
associated with this symmetry breaking is called the Majoron particle, 
$J$.

In principle,  Majoron particles can interact with matter 
--electrons, 
nuclei and photons. However the  cooling of red giant stars  
provides a strong 
bound on these interactions \cite{dearborn}.
Hereafter, we will assume that Majorons can  interact only  with 
neutrinos.
In the literature, two types of Majoron interaction have been studied:
\be \label{ainteraction}
{\cal L}_{int}={1 \over 2}J (g_{\alpha \beta} \Phi^T_\alpha \sigma_2 
\Phi_\beta+g_{\alpha 
\beta}^* \Phi^\dagger_\beta 
\sigma_2 \Phi_\alpha^*)
\ee
and
\be
\label{binteraction} 
{\cal L}_{int}=h_{\alpha \beta} \Phi_\alpha^\dagger  \bar 
\sigma.(\partial J) \Phi_\beta,
\ee 
where $J$ is the Majoron field,  $\Phi_\beta$ is 
a 
two-component representation of a neutrino of flavor $\beta$, $g_{\alpha 
\beta}$ and 
$h_{\alpha \beta}$ are $3\times 3$ coupling matrices. The  matrix 
$h_{\alpha \beta}$ is 
Hermitian while $g_{\alpha \beta}$ is a symmetric matrix. In the model 
\cite{peccei},  
 for a range of parameters, the interactions can be described by
 Eq. (\ref{ainteraction}) (see the 
appendix of  Ref. \cite{santamaria}). In this paper, we 
will use 
this  form of the interaction however, as we will see later, in  most 
cases our 
results apply for both forms.
Also, we will not assume any special condition on the diagonal or 
off-diagonal elements of $g_{\alpha 
\beta}$. Since we have chosen a general approach, our results apply 
to any massless scalar field that has an interaction of the form given by  
Eq. (\ref{ainteraction}), independent of the underlying model for it.

Majoron models are also interesting from the astrophysical point of view 
because, they provide the only mechanism for fast neutrino 
decay which has not yet been
excluded  (\cite{pakvasa}  and 
the references therein).
The role of neutrino decay in the solar \cite{solardecay} and 
atmospheric \cite{atmdecay} neutrino fluxes  has been 
extensively studied. The possibility of explaining the 
anomalies by pure neutrino decay is excluded.
In  Ref. \cite{beacom}, decay of solar neutrinos along with 
oscillation has been discussed and, for a normal hierarchical 
neutrino mass scheme, it has been found that
$$ |g_{21}|^2=|\sum_{\alpha,\beta} g_{\alpha\beta} U_{\alpha 
2}^* 
U_{\beta 1}|^2 \stackrel {<}{-}3 \times 
10^{-5}\left({10^{-5}\ \ {\rm eV}^2 \over \Delta m_{sun}^2}
\right).$$ 

In  Ref. 
\cite{long}, the different aspects and 
consequences of decay of neutrinos emitted by supernovae, have been 
studied. Future supernova 
observations can 
provide strong bounds on (or evidence for) neutrino decay and 
consequently on the Majoron coupling, provided that the 
uncertainties in supernova models are resolved.

If Majorons are coupled to neutrinos strongly enough, they can show up
in $ \beta \beta$-decay experiments, changing the spectrum of the final 
electrons. Non-observation of such an 
effect 
imposes a strong bound on the coupling constant 
\cite{betabeta}:
$$ |g_{ee}|<3 \times 10^{-5}.$$
 
Also, no sign of Majoron particles has been observed in the pion and kaon 
decays 
and therefore \cite{pika}:
$$ \sum_{l=e, \mu , \tau} |g_{el}|^2 \leq 3 \times 
10^{-5} \ 
\ \ \ {\rm 
and} \ \ \ \   \sum_{l=e, \mu , \tau} |g_{\mu l}|^2 
\leq 2.4 
\times 10^{-4}.
$$

The strongest bounds on neutrino-Majoron coupling are obtained by 
studying the role of these particles in a supernova explosion. 
In fact,
three types of bounds are obtained:
\newline
i) Deleptonization:
if the coupling, $|g_{ee}|$ is too large, Majoron 
emission can 
reduce the lepton number of the core of supernova via $\nu_e \to \bar 
\nu_e J$, preventing the emission 
of an intense
observable photon flux. In  \cite{smirnov,valle,tubbs,fuller} 
this 
effect has been studied; the result is
$$|g_{ee}|<2 \times 10^{-6}.$$
This bound strongly depends on the details of the supernova explosion 
model.
\newline
ii) Spectrum distortion: the production and absorption of 
the Majoron particle can 
affect the spectrum of the observed neutrino flux from a supernova 
explosion.  This effect has been studied 
in Refs. \cite{fuller,valle} and the result is 
$$ |g_{11}|=|\sum_{\alpha ,\beta} U_{\alpha 1}^* U_{\beta 1} 
g_{\alpha 
\beta}|<10^{-4}.$$
This result suffers from the low statistics of the SN1987a data and can be 
improved by future supernova observations.
\newline
iii) Energy loss: according to  \cite{binding}, the 
binding energy of a supernova core is  $E_b=(1.5-4.5)\times 10^{53}$ erg, 
which 
coincides very 
well with the energy emitted by SN1987a in 1-10 sec in the 
form of neutrinos.
Hence the power carried away by any exotic particle such as Majoron 
cannot be larger than $\sim 10^{53}$ erg/sec. This imposes strong bounds 
on the coupling of  Majorons. The 
effect of energy transfer due 
to Majoron emission has been studied in a number of papers 
\cite{smirnov,valle, santamaria, lam, peris, khol}. 

In the presence of  matter effects, a 
number of three-point processes that are kinematically forbidden in vacuum
become allowed. For example, neutrino decay becomes possible even in the 
absence of neutrino masses. Also, neutrino annihilation into a  
massless Majoron, $\nu \nu \to J$, becomes kinematically allowed.
The latter process has not been taken into account in  previous 
studies.
We will see that this is actually the dominant process contributing to 
energy 
loss in a supernova explosion. Previous studies must be reconsidered 
to  
take this effect into account.

In addition, the previous papers either considered only $g_{ee}$ or 
studied the Majoron couplings collectively without attention to the 
interplay of diagonal and off-diagonal couplings.
 In this paper, we study the effect of Majoron emission in the cooling 
process of supernova core considering all the relevant processes.
We find that even for very small 
values of coupling,  interplay of different processes may
change the 
neutrino densities inside the supernova, evading the bounds that  would be 
valid without this effect.

If the couplings are larger than some ``lower" bounds, Majorons will be so 
strongly trapped inside the supernova that they cannot give rise to 
significant 
luminosity. Note that these ``lower" bounds should be much larger than the 
limits at which  Majorons start to become trapped. For such large values 
of 
coupling, Majoron production can completely change the density profile of 
the core by transferring energy between different layers and by changing 
lepton numbers. In this paper 
we  discuss Majoron decay and all other processes that prevent  
energy transfer by Majoron particles and  derive the limits on coupling 
constants above which the produced Majoron cannot leave the core without 
undergoing scattering or decay.  We do not attempt to calculate 
any ``lower" bound on the coupling constants, because for large values of 
couplings, 
the density distributions inside the core need to be recalculated.
However we evaluate four-point processes which become important for large 
values of 
coupling constants. In summary, there is an ``upper" bound on coupling 
below which the rate of Majoron production is so low that it cannot affect 
the evolution of supernovae. The values of coupling above 
the ``upper" bound up to a ``lower" bound are not allowed. 
However, the values of coupling above the ``lower" bound 
(which are also higher than the ``upper" bound) are not 
forbidden by supernova cooling considerations because for 
such values of coupling, Majorons cannot escape the core 
freely. The forbidden range is then between the ``upper" and ``lower" 
bounds. In 
figure (1), we illustrate all the bounds on $|g_{ee}|$  to clarify the 
meaning of the ``upper" and ``lower" bounds. The shaded area is 
excluded by the supernova cooling process.

This paper is organized as follows. In  section 2, we calculate the 
cross-section of the relevant processes. In  section 3, we briefly review 
 the characteristics of the core. In  section 4, we derive 
the 
bounds on the coupling constants and the values above which the produced 
Majoron will scatter before leaving the core. Conclusions 
 are presented in  section 5.

\section{Majoron interactions}

In this section we first introduce the Lagrangian. Then, in  
subsection 
2.1, we derive the formulae for the neutrino propagator and the dispersion 
relation in the presence of matter.
The interaction rates for different processes involving Majoron are 
derived 
in  subsection 2.2.

 In the presence of matter, the 
Lagrangian of neutrinos 
can be written 
in the two-component formalism 
as \be
\label{lagrange}
{\cal L}=\Phi^\dagger_\alpha (i\delta_{\alpha \beta}\bar\sigma.\partial 
-V_{\alpha \beta}) \Phi_\beta 
-\frac{m_{\alpha \beta}}{2}(\Phi_\alpha^T 
C \Phi_\beta-\Phi_\beta^\dagger C\Phi_\alpha^*),
\ee
where $C=i\sigma_2$, $\alpha$ and $\beta$ are flavor indices, $\bar 
\sigma=(1,-\vec{\sigma})$ and $m_{\alpha \beta}$ is the   symmetric 
Majorana mass matrix.
The term $ \Phi^\dagger_\alpha V_{\alpha \beta} \Phi_\beta$ represents the 
matter effect. This term has a preferred frame, the frame of the supernova.
In the flavor basis, V is a diagonal matrix;
$V=diag(V_e, V_\mu, V_\tau)$ with
\be \label{vis} 
V_{e}=V_N+V_C, \ \ \ \ \ \ \ V_{\mu}=V_{\tau }=V_N,
\ee
where
\be 
V_C=\sqrt{2}G_F n_B(Y_e+Y_{\nu_e}), \ \ \ \ \ \ \ V_N=
\sqrt{2}G_F n_B(-{1 \over 2} Y_n+Y_{\nu_e}),
\label{wise}
\ee
$Y_i=(n_i-\bar n_i)/n_B$ and $n_B$ is  the baryon density 
\cite{book}.
\footnote{It is 
shown in  Ref. \cite{alex} that, if the neutrinos present in a 
medium are  coherent superpositions of different flavor states, 
the off-diagonal elements of $V_{\alpha \beta}$ can be nonzero. However 
inside the inner core the densities of $\nu_e$ and $\nu_\mu$ 
are different and 
the densities of $\nu_\mu$ and $\nu_\tau$ are very low and equal to the 
densities of $\bar \nu_\mu$ and $\bar \nu_\tau$, so the 
off-diagonal 
terms vanish.}  
In Eq. (\ref{wise}), the $Y_{\nu_e}$-dependent terms are the result of 
neutrino-neutrino scattering. Since in the medium of  interest 
(supernova core) $n_{\nu_\mu}=n_{\bar \nu_\mu}$ and 
$n_{\nu_\tau}=n_{\bar \nu_\tau}$,\footnote{ In section 3, we will see that 
these equalities are only  approximately true \cite{horo}.} the 
corresponding $Y$ 
parameters vanish 
and have been omitted from Eq. (\ref{wise}). In Ref. \cite{bote} it is 
shown that due to loop effects the values of $V_\mu$ and $V_\tau$ are 
slightly different, however, the difference is negligible: 
$V_{\mu}-V_{\tau}\simeq 5 \times 10^{-5}V_e$ \cite{ahmad}. In a typical 
supernova core, 
$V_{\mu }$ and $V_{e}$ are of the order of 10 eV and 1 eV, respectively. 

For the interaction term, we invoke the form of Eq. 
(\ref{ainteraction}). 
But we note that the derivative form of the interaction in Eq. 
(\ref{binteraction}) can be rewritten using  the equations of motion 
 as $$ -i  h_{\alpha \beta}m_{\beta \gamma }\Phi_\alpha^\dagger C 
\Phi_\gamma^* J-i 
h_{\alpha \beta}m_{\gamma \alpha} 
\Phi_\gamma ^TC\Phi_\beta.$$
Thus, for processes in which  all of the 
involved states are 
on-shell (in particular,  neutrino and Majoron decay,  $\nu \nu \to 
J$ and $\nu J \to \nu$) the two forms of interactions give the same 
results with the
replacement
\be \label{replace} g_{\alpha \beta} \to 
(h_{\alpha \gamma}m_{\gamma \beta}+m_{\alpha \gamma}h_{\gamma \beta}^T).
\ee
 As we 
will 
see, 
the 
most important processes
involve only on-shell particles. Therefore all of the bounds in this 
paper, 
apply for both derivative and pseudo-scalar forms of the interaction.

Majoron is a Goldstone boson associated with the {\it exact} B$-$L 
symmetry so 
in vacuum it is massless. Inside the supernova core Majoron obtains a tiny 
effective mass, $m_{eff}$,  due to elastic scattering off the background 
neutrinos. 
It can be
shown that $m_{eff}^2\sim |g|^2 N_\nu /q$ where $q$ 
is the typical momentum of the particles involved. For 
the values of coupling constants of order of the upper bounds the 
effective mass of Majoron is negligible ($m_{eff}^2/q \ll V_{e}$). The 
effective mass can be considerable only if $|g| \stackrel {>} {\sim} 5 
\times 10^{-4}$.

\subsection{The  propagators and the dispersion relation}
After straightforward calculations, we find
\be
\label{propagator}
 \sum_{\sigma ,\gamma}[(\bar\sigma \cdot p-V_{\alpha})\delta_{\alpha 
\sigma}-{m_{\alpha \gamma}m_{\gamma \sigma} \over 
p\cdot \sigma+V_{\gamma}}]\langle \Phi_\sigma 
(p)
\Phi_\beta^\dagger(-p) \rangle =i \delta_{\alpha \beta},
\ee
\be
\langle \Phi_\alpha^*(p)\Phi_\beta^\dagger(-p)  \rangle =\sum_\gamma C 
{m_{\alpha \gamma} \over p\cdot \sigma +V_{\alpha}}\langle \Phi_\gamma(p) 
\Phi_\beta^\dagger(-p)\rangle
\ee

and
\be \langle \Phi_\alpha (p) \Phi_\beta^T(-p) \rangle=\sum_\gamma
m_{\beta \gamma} \langle \Phi_\alpha(p) \Phi_\gamma^\dagger(-p) \rangle
{-1 \over p\cdot\sigma+V_{\beta}}C,
\ee
where all of the subscripts $\alpha$, $\beta$, $\gamma$ and $\sigma$ 
denote  $ 
\{e, \mu, 
\tau\}$.
As we will see, the diagrams in which these propagators are involved are 
important mainly when $|p|\stackrel {<}{\sim} {V_{\alpha}}$ so the 
effect of 
$V_{\alpha}$ must be treated non-perturbatively.
If the mass scale of 
neutrinos is high $(m_\nu \gg \sqrt{\Delta m^2})$,  
the masses are quasi-degenerate; $m_{\alpha \beta}\simeq m_\nu 
\delta_{\alpha \beta}$.\footnote{A proposed 
Tritium decay experiment, KATRIN \cite{katrin}, may be able to 
determine 
the mass scale.}
 In this case the formulae are simpler:
\be \label{nine}
\langle \Phi_\alpha (p) \Phi_\beta^\dagger (-p)\rangle={ -i 
\delta_{\alpha \beta}\over 
m_\nu^2-(p\cdot \sigma+V_{\alpha})(p\cdot \bar 
\sigma-V_{\alpha})}(p\cdot \sigma+V_{\alpha}),
\ee
\be
\label{ten}
\langle \Phi_\alpha^* (p) \Phi_\beta^\dagger (-p)\rangle = C {-i 
m_\nu \delta_{\alpha \beta} 
\over m_\nu^2-(p\cdot \sigma+V_{\alpha})(p\cdot \bar \sigma-V_{\alpha})}
\ee
and
\be \label{eleven}
\langle \Phi_\alpha (p) \Phi_\beta^T (-p)\rangle={ i m_\nu 
\delta_{\alpha \beta}
\over m_\nu^2-(p\cdot \sigma+V_{\alpha})(p\cdot \bar \sigma-V_{\alpha})}C.
\ee
Now let us find the dispersion relation. 
The Lagrangian (\ref{lagrange})
yields
\be \Phi_\alpha(p)^\dagger (-p\cdot\bar \sigma -V_{\alpha})= 
\sum_\beta m_{\alpha \beta}\Phi_\beta^T(p)C
\ee
and
\be
(p\cdot\bar \sigma-V_{\alpha})\Phi_\alpha(p)=-\sum_\beta m_{\alpha \beta} 
C\Phi_\beta^*.
\ee
Expanding the states as
$$ 
\Phi_\alpha(p)=\sum_{h=-1,1}u_\alpha(h,p)a_\alpha(h,p)+v_\alpha^\dagger(h,p)
a_\alpha^\dagger(h,p)\ 
\ \ \ {\rm for \ \ which} \ \ \ \ \vec{\sigma}\cdot\vec{p} 
u_\alpha(h,p)=h |\vec{p}|u_\alpha(h,p)
$$ 
we  find
\be
v_\alpha(h,p)=\sum_\beta m_{\alpha \beta} u_\beta^T {C \over p^0-h |\vec
{p}|+V_{\alpha}},
\ee
\be
u_\alpha(h,p)={-1 \over p^0+h|\vec 
{p}|-V_{\alpha}}\sum_\beta m_{\alpha \beta}C v_\beta^T(h,p)
\ee
and
 \be
\label{uu}
(p^0+h|\vec {p}|
-V_{\alpha})u_\alpha (h,p)=\sum_{\beta \gamma} 
{m_{\alpha \beta}m_{\beta \gamma} \over 
(p^0-h|\vec{p}| + V_{\gamma})}u_\gamma (h,p).
\ee
To find the dispersion relation and energy eigenstates we should solve
Eq. (\ref{uu}). 
Note that the dispersion relation depends on helicity.

For $m^2/ p\ll V \ll p$, one can easily show that 
\be \label{disperse} p^0_\alpha\simeq p-h V_{\alpha} + \sum_\beta 
{m_{\alpha \beta}^2 
\over 
2 p}\ee
and that the mixing among the flavors is of the order of 
$m^2 / 2 p(V_{\beta}-V_{\alpha})\ll 1$ which can be 
neglected. 
\subsection{The relevant decays and interactions}
In this subsection we first discuss the processes that 
produce Majorons, then we study those that annihilate or 
scatter them.
For illustrative reasons, in the following discussions, we ignore mixing 
({\it i.e.,} off-diagonal terms in both  coupling and mass matrix) 
and we  denote coupling, mass and effective potential by $g$, $m$ 
and $V$, neglecting their flavor indices.
In the cases that generalization is not straightforward, we will 
discuss the relevant steps.
Before beginning the detailed analysis, we should discuss an
important  conceptual point. As we see in Eq. (\ref{disperse}) the 
dispersion relation for neutrinos inside supernova is different from 
that in vacuum and hence some  reactions that are kinematically forbidden 
in vacuum, can take place in the supernova core.
As we will see the decay $\bar \nu \rightarrow \nu+J$ 
 and the interaction $\nu \nu \rightarrow J$  (or $\nu \rightarrow \bar 
\nu 
J$  and $ \bar \nu \bar \nu \rightarrow J$ 
depending 
on the sign of $V$)   are kinematically allowed.

In addition to  these three-point interactions, there are other 
interactions that 
produce Majorons:
\begin{itemize}
\item
$\nu+\nu \rightarrow J+J$ and $\bar \nu +\bar \nu \rightarrow J+J$;
\item
$\nu +\bar \nu\rightarrow J+J$.
\end{itemize}
As we will see the effect of the four-point interactions is negligible.

\subsubsection{$\bar \nu \rightarrow \nu 
+J$  or 
$\nu \rightarrow \bar \nu +J$}

In medium, if  $V$ is negative (positive), the decay  $\bar \nu 
\rightarrow
\nu +J$ ($\nu\rightarrow \bar \nu +J$) is possible.  Let us suppose 
$V<0$, then, without loss of generality, 
we can
write $$ p_{\bar \nu}=(p_i-V,0,0,p_i) \ \ \ \ p_\nu=(p_f+V,p_f \sin
\theta,0,p_f \cos \theta)$$ where we have neglected corrections of order 
of 
$m^2/p_i \ll V$. Energy-momentum conservation implies that,  $$ 
p_J=(p_i-p_f-2V,-p_f \sin
\theta,0,p_i-p_f \cos \theta).$$ Recalling that we have neglected the 
effective mass of Majoron,
the process $\bar \nu \rightarrow
\nu +J$ is kinematically allowed if and  only if $p_J^2=0$
 and 
all the zeroth components of the four-momenta are positive.
$p_J^2=0$ implies 
\be \label{onshell1}1-\cos\theta={4 V^2-2 V (p_i-p_f) \over 
2 p_i p_f}.\ee 
For $|V|<p_f<p_i$, the above equation can be satisfied with all of the 
energies  positive. This means that 
the process is kinematically allowed.

Restoring flavor indices, it can be shown that for  $V_{ 
\beta}+V_{\alpha}<0$ the rate 
of  $\bar \nu_\alpha \rightarrow
\nu_\beta +J$ is given by
 \be {d \Gamma \over d p_f}={|g_{\alpha
\beta}|^2 \over 8 \pi}{p_i-p_f \over p_i^2}|V_{\alpha}
+V_{ \beta}|F^F_{\beta}(p_f)+{\cal
O}(\frac{m^2}{p^2}) \label{nudecays} \ee where $p_i$ and $p_f$ are
the momenta of the initial and final neutrinos and $p_f$ extends
from Max$(\frac{1}{2} |V_{ \alpha} +V_{\beta}|,- V_{
\beta})$ to $p_i$. In the equation, we have also 
included the Fermi factor $$F^F_{\beta}(p_f)=(1-{1\over 
e^{\frac{E-\mu}{T}}+1})$$ which reflects
the fact that inside the supernova some states have already been
occupied by neutrinos.

Similarly, for $V_{ \alpha}+V_{ \beta}>0$, 
the process $\nu_\alpha \rightarrow \bar \nu_\beta+J$ can take place. The 
decay rate is given by Eq. (\ref{nudecays}) replacing 
$F^F_\beta(p_f)$ with $F^F_{\bar\beta}(p_f)$.  The range of 
$p_f$ 
extends from Max$(\frac{1}{2}|V_{ \alpha} +V_{\beta}|,V_{\beta})$ to $p_i$. 

\subsubsection{ $\nu+ \nu \rightarrow J$ or
$\bar\nu +\bar\nu \rightarrow J$
}
In vacuum,  the processes $\nu+ \nu \rightarrow J$ or
$\bar\nu +\bar\nu \rightarrow J$ are not 
kinematically allowed. However in medium, where $V$ is 
negative (positive) the process  $\nu +\nu \rightarrow J$ ( 
$\bar \nu 
+\bar \nu \rightarrow J$) can occur.
Let us suppose $V<0$ and study the possibility of $\nu+\nu 
\rightarrow J$. Without loss of generality, we can write the 
four-momenta of the initial neutrinos as
$$ p_1=(p_1+V,0,0,p_1) \ \ \ \ \ \ p_2=(p_2+V,p_2 \sin 
\theta,0, p_2 \cos \theta),$$
for which $p_1+V$ and $p_2+V$ are both positive. Energy-momentum 
conservation implies
$$p_J=(p_1+p_2+2V,p_2 \sin \theta,0,p_1+p_2\cos \theta).$$
Recalling that we have neglected the effective mass of Majoron,
the process $\nu(p_1)+\nu(p_2) \rightarrow J(p_J)$
will be kinematically allowed if and only if $p_J^2=0$ 
or
\be \label{onshell2}1-\cos \theta =-{2 V^2+2V(p_1+p_2) \over 
p_1 p_2}\ee
which can be  satisfied for $p_1,p_2>|V|$.

Neglecting $V^2/p^2$ effects, for 
 $V_{ \alpha} + V_{ \beta}<0$, it can be shown 
that the cross section  of   $\nu_\alpha 
\nu_\beta
\rightarrow J$ is given by 
\be \label{barchasb}
\sigma={(2 \pi) |g_{\alpha \beta}|^2 \over 4 p_1^2 
p_2^2|v_1-v_2|}(p_1+p_2)|V_{\alpha}+V_{ 
\beta}|\delta (\cos \theta-\cos \theta_0)
\ee
where $p_1$ and $p_2$ are the momenta  of 
the two initial particles, $\theta$ is the angle 
between them and $\cos \theta_0=1-{|V_{\alpha}
+V_{ \beta}|(p_1+p_2) \over p_1 p_2}$.

Similarly, it can be shown that for $V_{ 
\alpha}+V_{\beta }>0$, instead of $\nu_\alpha+\nu_\beta 
\rightarrow J$, the process  $\bar \nu_\alpha+\bar \nu_\beta 
\rightarrow 
J$ can take place with the cross section again given by Eq. 
(\ref{barchasb}).

 \subsubsection{The process
$\nu+ \bar \nu \rightarrow J+J$}
For reasons that will become clear in a moment, we analyze $\nu$ and $\bar 
\nu$ as wave packets rather than as plane waves.
Let us ignore the neutrino mass for simplicity. Then, calculating diagram 
(c) in figure 2, we find 
\be \label{hassle}2\pi i{\cal M}=(2 \pi)^4 (i 
g) 
(2 \pi)^4 (i 
g^*){1 \over (2 \pi)^6} {1\over \sqrt{4k^0_1 
k^0_2}}\times \int \int f(p_2) u^T(p_2)\sigma_2 {1 \over (2 
\pi)^4}\times\ee
$$\left( {q^0-V-\vec{q}\cdot \vec{\sigma} \over (q^0-V)^2- 
|\vec{q}|^2+i\epsilon}-
2 \pi \{\theta(-q^0)+\epsilon(q^0)(1-F^F(q^0))\}\delta 
[(q^0)^2-(q+V)^2]\right)$$
$$
\times \sigma_2 \nu
(p_1)\bar f(p_1)d^3p_1 d^3p_2+(k_1\leftrightarrow  k_2),$$
where $k_1$ and $k_2$ are the momenta of  the Majorons and 
$\int f (p_2) |p_2 \rangle d^3p_2$ and $\int \bar f 
(p_1)|p_1 
\rangle  d^3p_1$
 represent the states of the neutrino and anti-neutrino, 
respectively. 
In  Eq. (\ref{hassle}), $q=k_2-p_2$ and we have considered the matter 
effects in the propagator: $$F^F(q^0)=1-1/[exp((q^0-\mu)/T)+1]$$ is the 
Fermi factor. 

For
both  positive and negative $V$, in 
the vicinity of $(\vec{k_1}=\vec{p_1},\vec{k_2}=\vec{p_2})$ and  
$(\vec{k_1}=\vec{p_2},\vec{k_2}=\vec{p_1})$, there are poles which are  
non-integrable 
singularities. Without the wave packets, the 
total cross section would be divergent. Setting $m_\nu$ non-zero just 
shifts the 
pole a little bit and does  not solve this 
problem. This is due to the fact that for negative 
(positive) $V$, the processes $\bar \nu \rightarrow 
\nu+J$ and $\nu+ \nu \rightarrow J$ ( 
$ \nu \rightarrow\bar \nu +J$ 
and $\bar \nu+ \bar \nu \rightarrow J$)
 can take place on shell, so the singularity is indeed a physical one.
Essentially, for $V<0$ the reaction $\nu \bar \nu \rightarrow JJ$ can 
proceed in two 
steps, first, $\bar \nu \rightarrow \nu J$ and later,
at a completely distant place $\nu \nu \rightarrow J$.
In other words, the total cross section has two parts: i) a ``connected" 
part; ii) a ``disconnected" part which can be considered as two successive 
three-point processes.

Let us now consider in more detail the relation between Eq. 
(\ref{hassle}) and its component three-point processes.
  For definiteness, we consider the case  $V<0$.
We have explained that the reaction $\bar \nu \nu \to J J$ contains a 
subprocess that factorizes as
$$ \int_q\langle J_1 J_2|
\nu \nu(q) 
J_1\rangle \langle
\nu(q) 
J_1 |\bar \nu \rangle .$$ 
More explicitly, the factorized amplitude takes the form
\be \label{catch}
2\pi i{\cal M}=
 {-g g^* \over (2 
\pi)^6\sqrt{4k^0_1 k^0_2} }\int \int \int \int
\int_\tau^{2\tau}\int_{-\tau}^\tau f(p_2) u(p_2)e^{((\vec 
{p}_2-\vec{k}_2)\cdot\vec{x}_2-(p^0_2-k^0_2)x^0_2)} 
\ee
$$
 \sigma_2{1 \over (2 \pi)^3} 
\int 
{q^0-V-\vec{\sigma}\cdot\vec {q} \over 2 
|\vec{q}|}F^F(q^0)e^{-i 
q^0(x^0_2-x^0_1)}e^{i\vec{q}.(\vec{x}_2-\vec{x}_1)}\delta (q^0-V 
-|\vec{q}|)
d^4q)$$ $$\sigma_2 v 
(p_1)\bar f(p_1) 
e^{i((\vec{p}_1-\vec{k}_1).
\vec 
{x}_1-(p^0_1-k^0_1)x^0_1)}d^3 p_1 d^3p_2 dx^0_1 
dx^0_2 d^3x_1 d^3x_2, $$
where $F^F(q^0)=1-1/(exp((q^0-\mu)/T)+1)$ represents the matter effects.
In  Eq. (\ref{catch}), $\tau$ represents the boundaries on 
time integrations and therefore it must be very large ({\it 
i.e.,} 
$\tau \stackrel {>}{\sim} 5/V$). We have 
written the time boundaries explicitly to emphasize the 
causality conditions.
Transferring the amplitude for $\bar \nu \nu \rightarrow 
JJ$ (Eq. (\ref{hassle})) from momenta $p_1$, $p_2$ to coordinates $x_1$, 
$x_2$, it can be shown that for the region $|x_2-x_1|>2 \tau 
\stackrel 
{>}{\sim} 10/V$, these two correspond.
Therefore, if the initial neutrino and anti-neutrino are localized 
at  distance $R> 10/V$, their interaction 
rate can be calculated  by Eq. (\ref{catch}) instead of Eq. 
(\ref{hassle}).

Consider  $\nu$ and $\bar 
\nu$ which are localized  at distance  $R> 
10/ V$ far from each other.
We have shown that their interaction cross section is given by  
$ |\langle J_1 
J_2| \nu \nu
J_1\rangle \langle
\nu
J_1 |\bar \nu \rangle |^2$. So, this interaction can be considered as two 
subsequent processes. First $\bar \nu$ decays into $J_1$ and $\nu$. Then, 
the produced neutrino propagates a distance $R$ and annihilates with the 
other $\nu$ into $J_2$.
 In other words, to calculate the interaction 
probability  of  two such states, we can consider 
$\bar \nu \rightarrow \nu J$ as an 
additional source for $\nu$ and 
consequently the process $\nu \nu 
\rightarrow J$. This can be compared to the more familiar sources of 
neutrinos like electron capture, $ \langle J |\nu \nu \rangle \langle \nu 
n 
|e^- p^+ \rangle $.  

Of course the set of states that  are localized at distance
$R>10/V\sim 10^{-9} R_{core}$   far from one another is not a complete 
set. We should
also consider the states which are closer and/or have overlap 
with each other. 
If we rewrite   Eq. (\ref{hassle}) in the x-coordinates, as we have done 
in Eq. (\ref{catch}), calculation of the amplitude of two states localized 
next to each other at  distance $R$ will be easier.  For such two states, 
the integral for 
$|\vec{x_1}-\vec{x_2}|>10/V$ vanishes (because of 
the specific form of $f(p_1)$ and $\bar f(p_2)$), so 
  we can restrict the integration over $|\vec{x_1}-\vec{x_2}|$ to the interval (0, $10/V$).
For $|q^0-V-|\vec{q}|| \gg |V|/10$, the amplitude for two 
states 
localized at $R<10/V$ far from each other is equal to the amplitude for 
states with definite momenta, but for 
$|q^0-V-|\vec{q}||<|V|/10$, 
the amplitude for the two localized states is much smaller. 
This is because in calculation of amplitude for two states with definite  
momenta, we encounter an integration $\int_0^\infty g(x) 
e^{i(q^0-V-|\vec{q}|)x} dx$ which diverges for 
$q^0-V-|\vec{q} |\rightarrow 0$ but for two states which are 
localized next to each other the corresponding integration is 
$\int_0^{10/V} g(x)e^{i(q^0-V-|\vec{q}|)x} dx$ which is finite. 
The total cross section for neutrinos and anti-neutrinos 
localized 
next to each other is then given by 
an angular integral over the square of (\ref{hassle}) in which
the integration is 
over all angles except those for which 
$|q^0-V-|\vec{q}||<|V|/10$.
Consider the special case that the sum of the momenta of $\nu$ 
and $\bar \nu$ is zero. Setting the cutoff equal to $\lambda |V|/10$, 
($\lambda$ is an arbitrary number of order of one)
for such two particles 
we obtain \be \label{tot}\sigma_{tot}\sim {|g|^4 
\over 
8 \pi p_1 p_2 |v_1-v_2|}\left[ \ln  \left(\frac {p_1 
p_2}{(\lambda V/10)^2}\right)+{10\over \lambda}-{14 \over 4}\right] 
. \ee
Since we have a preferred frame (the frame of the supernova), 
the total cross section is not Lorentz invariant and for 
each $\nu$ and $\bar \nu$ the total cross section must be 
calculated, independently. Now consider a pair of a neutrino 
and an anti-neutrino that make a general angle. Then Eqs. 
(\ref{onshell1},\ref{onshell2}) show that in the vicinity of
singularity the momentum flowing in the propagator is of the 
order of $|V|$ ({\it i.e.,} $|\vec{q}|=|\vec{p}_2-\vec 
{k}_2|\sim |V| $). Therefore for scattering angles that 
$(q^0-V-\vec {q} \cdot \vec {\sigma} )/\left[ 
(q^0-V)^2-|\vec 
{q}|^2\right] \sim 1/V$, the phase factor $(\int d^3k_1d^3k_2 
\delta^4(p_1+p_2-k_2-k_1))$ is of the order of $|V|^2/p^2$. Thus 
the total cross section has no strong dependence on $|V|$ 
and for general initial momenta the cross section can be 
estimated by Eq. (\ref{tot}).
 
Here, for simplicity we have dropped the flavor indices but 
for the more
general case the discussion is similar.


\subsubsection{The process $\nu+ \nu 
\rightarrow J+ J$ and $\bar \nu+\bar \nu\rightarrow J+ J$ }
The discussion of $\nu \nu \to JJ$ and $\bar \nu+\bar \nu\rightarrow J+ J$ 
can be carried out 
in a similar way.
For quasi-degenerate neutrino masses, the amplitude for $\nu \nu     
\rightarrow J J$ (see diagram (d) in Fig. (2)) is given by,
\be \label{nunujj} \frac{1} {(2\pi)^2\sqrt{4 k^0_1 k^0_2}}\sum_\gamma \int 
\int f_1(p_1)     
u_\alpha^T (p_2) C (ig_{\alpha \gamma})(ig_{\gamma \beta})\times \ee
 $${i m 
(m^2+V_\gamma^2+q^2-q_0^2+2 \vec{q}\cdot \vec{\sigma} V_\gamma)  
\over
(m^2-q_0^2+(V_\gamma-q)^2)(m^2-q_0^2+(V_\gamma+q)^2)}\times
$$
$$ f_2(p_2)u(p_2) d^3p_1 d^3p_2 + (k_1 \leftrightarrow k_2)+{\cal A},$$
where $\int f_1 (p_1) |p_1 \rangle d^3p_1$ and $\int  f_2(p_2)|p_2 \rangle  
d^3p_2 $ represent the 
initial neutrino states, $k_1$ and $k_2$ are the momenta of the emitted 
Majorons and $q=k_2-p_2$.
 The term ${\cal A}$ summarizes all of the Fermi effects on the propagator. 
The  amplitude  for values of $q$ which 
$(q+V_\gamma)^2-q_0^2-m^2 \sim p_1^2, p_2^2 \gg m^2, 
V_\gamma^2$ is 
negligible, and the main contribution to the cross section comes from 
the small 
solid angle ($\sim V^2/p_1p_2$)  for which 
$(q+V)^2-q_0^2-m^2 
\stackrel {<}{\sim} V^2$.

First, let us discuss the process $\nu_e \nu_e \to JJ$. In general, for 
$\gamma=\mu , 
\tau$, there are singularities which  
correspond to an on-shell $\nu_\mu$ 
or $\nu_\tau$. 
Note that, if $\vec{p_1}$ and $\vec {p_2}$ are parallel or make an angle 
smaller than $\sim |V_{e}/V_{\mu}|$, the singularities disappear. As for 
the case $\bar \nu \nu \to JJ$, we can discuss 
that if the 
initial states are localized at distance $R> 10/V_\mu$ far from each 
other, the process $\nu \nu \to JJ$ will be equivalent to two successive 
processes $\langle \nu_{\mu (\tau)} J | \nu_e \rangle$ and then $\langle 
J|\nu_e \nu_{\mu (\tau)} \rangle$.
This yields a cutoff of $V_\mu/10$ for calculating the 4-point total 
cross-section. Note that although $\nu_e \to J+\nu_\mu$ is kinematically 
allowed ($V_\mu<V_e$), $\Gamma (\nu_e \to J \nu_\mu)$ is suppressed by 
$(m/p_{\nu_e})^2$ and in practice, will not have any significant effect.

For $\gamma=e$, there is no singularity (except for the case that one of 
the Majorons is soft and  $\nu_e \nu_e \to J$ is 
kinematically possible) and therefore no cutoff is needed.
The total cross section can be estimated as 
\be \sigma_{tot}(\nu_e \nu_e \to JJ)={1\over |v_1-v_2|(2 \pi)^2 p_1 p_2} 
\left(a 
|g_{e \mu}^2+g_{e \tau}^2|^2 
(\frac{m}{V_\mu})(\frac{m}{V_\mu/10})+b  |g_{ee}|^4 
\right). \ee
Similarly,
\be  \sigma_{tot}(\nu_e \nu_{\mu(\tau)}) \to JJ)={1\over |v_1-v_2|(2 
\pi)^2 p_1 
p_2}\times \ee $$
\left( a'|g_{e \mu}g_{\mu \mu(\tau)}+g_{e \tau}g_{\tau \mu (\tau)}|^2
(\frac{m}{V_\mu})(\frac{m}{V_\mu/10})+b' |g_{ee}g_{e \mu 
(\tau)}|^2\right)
$$
and 
\be \sigma_{tot}(\nu_{\mu(\tau)} \nu_{\mu(\tau)}\to JJ)={1\over 
|v_1-v_2|(2 \pi)^2 
p_1
p_2}\left(a''|g_{\mu \mu(\tau)}^2+g_{\tau \mu(\tau)}^2|^2+b''|g_{e \mu 
(\tau)}|^4\right).
\ee
with $b$, $b'$, $b''$,
$a$, $a'$ and $a''$   of order  1.
In  Ref. \cite{peris}, $  \sigma_{tot}(\nu_e \nu_e \to JJ)$ has been 
calculated, 
ignoring $V$ and the off-diagonal elements of the coupling matrix. The 
result agrees with our 
estimation in the sense that the term proportional to $|g_{ee}|^4$ is not 
suppressed
by $m$.

The total cross section for $(\bar \nu_\alpha \bar\nu_\beta \to JJ)$ 
is equal to $\sigma_{tot}( \nu_\alpha  \nu_\beta \rightarrow JJ)$ 
replacing
$V$ with $(-V)$.

\subsubsection{The processes $\nu +J \rightarrow \bar \nu$ or 
$\bar \nu +J \rightarrow \nu$:}

These processes are the opposite of anti-neutrino and neutrino decay and, 
hence, the kinematical conditions are similar.
If $V_{ \alpha}+V_{ \beta}$ is negative 
(positive) the process  $\nu_\alpha J \rightarrow \bar 
\nu_\beta$ (  $\bar \nu_\alpha J \rightarrow 
\nu_\beta$) can take place with cross section
\be \label{jannihilates}
\sigma={(2 \pi)\over 4p q |v_1-v_2|} |g_{\alpha 
\beta}|^2 {|V_{ \alpha}+V_{ \beta}| \over p}
F^F_{\stackrel {(-)} {\beta}}(p+q)\delta (\cos 
\theta -\cos 
\theta_0)
\ee
where $p$ and $q$ are the momenta of the initial 
neutrino and Majoron, respectively. 
$\theta$ is the angle between the two initial states and 
$$\cos \theta_0=-1+{(p+q)|V_{ \alpha}+V_{ 
\beta}| \over p q}.$$
$F^F_{\stackrel {(-)} {\beta}}(p+q)$ is the Fermi 
factor for the final state.

\subsubsection{The Majoron decay, $J \rightarrow \nu +\nu$ {\bf or} 
$J\rightarrow 
\bar \nu+ \bar \nu$}

The decay $J\rightarrow \nu \nu$ ($J \rightarrow \bar \nu \bar \nu$)
is the opposite of the interaction $\nu \nu \rightarrow J $ ($\bar \nu 
\bar 
\nu \rightarrow J$) and therefore the kinematics are similar.

For $V_{\alpha}+V_{ \beta}<0$ ($V_{ 
\alpha}+V_{\beta }>0$) the Majoron can decay into 
$\nu_\alpha +\nu_\beta$ ($\bar \nu_\alpha +\bar 
\nu_\beta$) and the decay rate up to a $(|V|/p_i)^2$ 
correction is given by 
\be \label{Jdecay}
d\Gamma ={|g_{\alpha \beta}|^2 \over 8 \pi}{|V_{ 
\alpha}+V_{ \beta}| \over p_i}\int_{0}^{p_i} 
F^F_\alpha (p_f)F^F_\beta (p_i-p_f)dp_f
\ee
where $p_i$ and $p_f$ are the momenta of the Majoron  
and either of the final neutrinos, respectively. $F^F_\alpha$ 
and $F^F_\beta$ are the Fermi factors reflecting the fact 
that in the core of the supernova some states have been 
occupied by 
already present neutrinos.  

\subsubsection{The processes $\nu+ J \to \bar \nu +J$  and $\bar \nu +J\to 
\nu + J$}
The amplitude for $\nu_e +J \to \bar \nu_e +J$ has two singularities in 
the 
$t$-channel due to $\nu_\mu$ exchange.
Using a similar discussion to that  in section 2.2.4, it can be 
shown 
that these  singularities may be considered as two successive 
three-point interactions $ \langle \bar \nu_e |J \nu_\mu \rangle \langle 
\nu_\mu J 
|\nu_e \rangle$ and $\langle J|\nu_e \nu_\mu \rangle \langle \bar \nu_e 
\nu_\mu |J \rangle$. This yields a cutoff $\sim |V_\mu|/10$ around the 
singularity to determine the four-point interaction. 
In the case of head-on collision where the initial  particles are within
a small solid angle $\sim (V/p)^2\ll 4 \pi$ around $\cos \theta=-1$, there 
will be 
another singularity in the $s$-channel which can be considered as $\langle 
\bar \nu_e J| \bar \nu_\mu \rangle \langle \bar \nu_\mu |\nu_e J
\rangle$.
We recall that any
discussion about $\nu_\mu$  applies  to
 $\nu_\tau$ as well, because these states are completely equivalent for 
the
supernova evolution.
The total cross-section for $\nu_e J\to \bar \nu_e J$ can be evaluated as
\be {1 \over (2 \pi)^2|v_1-v_2|p_1p_2}\left( a|g_{e \mu}^2+g_{e 
\tau}^2|^2(\frac{m^2}{V_\mu^2/10}) +b|g_{ee}|^4 \right)F^F ,
\label{form}
\ee
where $a\sim b \sim 1$ and $F^F$ is the Fermi-blocking factor for the 
final 
neutrino.
A similar discussion holds for $\bar \nu_e J\to \nu_e J$, and the 
corresponding cross-section is also of the form of Eq. (\ref{form}).

The processes $\nu_\mu +J \to \bar \nu_e+J$, 
 $\nu_e +J \to \bar \nu_\mu+J$,  $\bar \nu_\mu +J \to  \nu_e+J$ and  
$\bar \nu_e +J \to  \nu_\mu+J$ also have  singularities in the $t$-channel 
due 
to $\nu_\mu$-exchange and can be considered as two successive 
three-point processes. Following the same discussion as in  sections 
2.2.3 and 2.2.4, we use the cutoff $\sim V_\mu/10$ to evaluate the cross 
section for the four-point interactions. The cross-sections of these 
processes have the form
\be {1 \over (2 \pi)^2|v_1-v_2|p_1p_2}\left( a|g_{e \mu}g_{\mu \mu}+g_{e
\tau} g_{\mu \mu}|^2(\frac{m^2}{V_\mu^2/10}) +b|g_{ee}g_{e \mu}|^2 
\right)F^F 
,
\label{format}
\ee
where $a\sim b \sim 1$ and  $F^F$ is 
the Fermi-blocking factor for the
final neutrinos. 
The processes $\nu_\mu J \to \bar \nu_e J$
and $\bar \nu_e J \to \nu_\mu J$ can also have  singularities in the 
$s$-channel 
only 
if the initial particles are almost parallel, {\it i.e.}, if their 
relative 
angle resides 
within a small solid angle $\sim (V/p)^2 \ll 4 \pi$ around $180^\circ$. 
We can safely neglect such states.

For the process $\nu_\mu J \to \nu_\mu J$, there is no singularity and it 
is straightforward to show that the cross section is of the form,
\be {1 \over (2 \pi)^2|v_1-v_2|p_1p_2}\left( a|g_{\mu \mu}^2+g_{\mu 
\tau}^2|+b|g_{\mu e}|^2 \right).
\ee


\subsubsection{The processes $\nu+J \rightarrow \nu +J$ and 
$\bar\nu+J \rightarrow \bar\nu +J$} 
In general, the process $\nu +J \rightarrow \nu +J$ has a 
singularity in the 
$t$-channel. 
With a similar 
discussion as in  section 2.2.3, we can show that this singularity 
can be evaluated as two successive three-point interactions $\langle 
J | \nu \nu \rangle \langle \nu \nu |J \rangle$ resulting in a cutoff 
of the order of $V/10$ for evaluation of the four-point interactions. 
Using 
this cutoff, the cross section is of the
order of \be {|g|^4 \over (2 \pi)^5|v_1-v_2|p_1 p_2} \ln \left( \frac{p_1 
p_2}{V^2/100}\right) F^F,
\label{shekl}
\ee  
where $F^F$ is the Fermi-blocking factor for the
final neutrino. 

If the initial particles undergo a  head-on collision ({\it i.e.,} they 
are within
a small solid angle $\sim (V/p)^2$ around $180^\circ$) there will be 
another singularity in the $s$-channel which can be considered as $\langle 
\nu J| \bar \nu \rangle \langle \bar \nu |\nu J
\rangle$.
The process $\bar \nu +J \to \bar \nu+J$ has one singularity
which can be evaluated as $\langle \bar \nu |J \nu \rangle \langle J \nu | 
\bar \nu \rangle$. Again the cross section is of the form of Eq. 
(\ref{shekl}). 
\section{Supernova core without Majorons}
The dynamics of a supernova explosion is described in a number of articles 
and books 
({\it  e.g.,}  \cite{book}).
Here we only review the aspects of the supernova explosion which are 
relevant 
for 
our calculations.
\newline
Very massive stars (${\cal M}> 8{\cal M}_\odot$), at the 
end of their 
lifetime,
develop a degenerate core with a mass around
$1.5 {\cal M}_\odot$ made up of iron-group elements.
 As the outer layer 
burns, it deposits more iron that 
adds to the mass of the core.  Eventually the core reaches its 
Chandrasekhar limit,  at which the Fermi-pressure of the electron gas 
inside the 
core cannot support the gravitational pressure, and 
the star collapses. The 
collapse forces  nuclei to absorb the electrons via 
$e^-+p\rightarrow n+\nu_e$.
At the early stages, the produced $\nu_e$ can escape from the core 
but, eventually, the core becomes so dense 
that even neutrinos are  trapped. 
The layer beyond which neutrinos can escape without 
scattering  is called the ``neutrino-sphere". 

As the density of the central core reaches nuclear density ($\rho\simeq 3 
\times 10^{14} \ \ \ {\rm g}/{\rm cm}^3$), a shock wave builds up which 
propagates outwards. We will refer to the pre-shock stage as the infall 
stage.
This stage takes only around 0.1 sec.
As the shock wave reaches  the neutrino-sphere, it dissociates the heavy 
nuclei. The dissociation has three different results:
\begin{enumerate}
\item
It consumes the energy of the shock, so that the shock eventually stalls;
\item
It allows   neutrinos to escape more easily;
\item
 It liberates protons that interact with the electrons present in the star
($e^-+p\rightarrow n+\nu_e$),  giving rise to the famous ``prompt $\nu_e$ 
burst". The prompt $\nu_e$ 
burst deleptonizes the star but carries only a few percent of the total 
energy.
\end{enumerate}
The stalled shock should regain its energy. Otherwise, it cannot 
propagate further and give rise to the spectacular fireworks. 
According to the models, this energy is provided by $\nu_e$ diffusing 
from the inner core to outside. The density of $\nu_e$ inside the inner 
core is very high. The 
corresponding Fermi energy is $\sim 200$ MeV while the temperature is only
around 10 MeV. 
At the beginning the temperature of the neutrino-sphere is around 20 MeV.
So the diffused neutrinos leave their energy as they travel outside,  
warming up the core. This energy can revive the shock. (In 
fact, this 
mechanism is controversial \cite{book}, but we will not use 
the shock revival 
mechanism for our calculations. 
Most of our calculations are related to the inner core, which is free of 
these controversies.)
The temperature in the 
outer core increases to 40 MeV; actually, the outer core and the 
neutrino-sphere become warmer than the center. At the outer core, 
neutrinos of each type ($\nu_e$, $\bar \nu_e$, $\nu_\mu$, $\bar \nu_\mu$,
$\nu_\tau$ and  $\bar \nu_\tau$) are present. These neutrinos 
escape the star and deplete its binding energy ($E_b=(1.5-4.5)\times 
10^{53}$ 
erg \cite{binding}).

Two kinds of ``upper" bounds can be imposed on the neutrino-Majoron 
couplings by studying supernova evolution:
\newline
1) If the coupling constant is too large, the process $\nu_e \rightarrow 
J+ \bar \nu_e$, during the infall stage, deleptonizes the core and 
according to the models \cite{some} a
successful explosion cannot occur. This bound has been correctly studied 
in 
\cite{smirnov,valle} and the result is $g_{ee}\stackrel {<}{\sim} 2\times 
10^{-6}$.
\newline
2) If the coupling is non-zero, Majorons can be produced inside the inner 
core and can escape freely from the star, depleting the binding energy.
The observed neutrino pulse from SN1987a coincides with that predicted 
by  current 
supernova models. This means that the energy carried away by Majorons
(or any other exotic particles)
should be smaller than the binding energy. 
The Majoron luminosity, ${\cal L}_J$, as large as $10^{53}$ erg/sec could 
significantly affect the neutrino pulse. Here, we will take ${\cal 
L}_J<3\times 10^{53}$ erg/sec as a conservative maximum allowed value.   
This gives  an 
upper bound on the coupling constants.

If the coupling of Majorons is larger than a ``lower bound", the Majorons 
will be 
trapped so strongly that their luminosity will be small. We will discuss 
this case later.

Let us review the characteristics of the 
core. The inner core ($R<R_{inner}\sim 10$ km)  to a 
good approximation is homogeneous. The density in the inner core is 
around $5 \times 10^{14}$ g/cm$^3$.
The distributions of all types of neutrinos follow the Fermi-Dirac formula 
with $T_{inner}\sim 10-30$ MeV and different chemical potentials 
\cite{adam,pons}.
As mentioned earlier, the chemical potential for $\nu_e$ is 
around 200 
MeV. So, inside the inner core, $\nu_e$ is degenerate while 
the density of $\bar \nu_e$ is negligible ($\mu_{\bar 
\nu_e}=-\mu_{\nu_e}=-200$ MeV). 
The suppression of the density of $\bar \nu_e$ is due to absorption on 
electrons. 
In the first approximation, the chemical potentials for 
$\stackrel {(-)}{\nu_\mu}$ and  
$\stackrel 
{(-)}{\nu_\tau}$ are equal to zero. 
In 
Ref. \cite{horo}, it is shown that, because the interactions of 
$\nu_\mu$ and $\nu_\tau$ with matter are slightly stronger 
than the interactions of $\bar \nu_\mu$ and $\bar \nu_\tau$, 
their chemical potentials become nonzero: 
$\mu_{\nu_\mu}/T=\mu_{\nu_\tau}/T\simeq 5T/m_p<1$. We 
will neglect 
$\mu_{\nu_\mu}$ and $ \mu_{\nu_\tau}$ in our analysis. In fact 
the large uncertainty in the determination of the temperature 
affects our results more dramatically. 
The presence of $\mu$ in supernova can break the equivalence of $\nu_\mu$ 
and $\nu_\tau$. However, we neglect this effect and treat $\nu_\mu$ and 
$\nu_\tau$ in exactly  the same way. In  Table 1, we show the values of 
$V_{e}$ and $V_{ 
\mu}$ $(=V_{\tau })$ at  different instants after the bounce inside the 
inner core. The values of $Y_e$ and $Y_{\nu_e}$ are taken from  Ref. 
\cite{adam}.

 \begin{table} 
\begin{center}
\renewcommand{\baselinestretch}{1.0}
\renewcommand{\baselinestretch}{2.0}
\label{limits4}

\begin{tabular}{|c|c|c|} \hline
t(sec)& $V_{e}$ \ \ (eV)& $V_{ \mu}=V_{ \tau}$ \ \  (eV) \\
\hline
0 & 2.3 & -11.7 \\
0.5 & 1 & -12.3 \\ 
1 & -0.3 & -12.8 \\
1.5&-1& -13.1 \\
\hline
\end{tabular}
\baselineskip 0.1 cm
\caption{The values of the effective potentials at different instants 
after bounce without Majoron production.}
\end{center}
\end{table}
Outside the inner core, $ R_{inner}\sim 10 \ \ {\rm km}<R<R_{out} \sim 15 
\ \ {\rm km}$, the density of $\nu_e$ is much lower, 
$\mu_{\nu_e}/T 
\stackrel {<}{\sim} 1$, but instead the density of $\bar \nu_e$ is higher 
than in the inner core. In fact, in the outer core 
($R_{inner}<R<R_{out}$), thermal equilibrium for neutrinos is only an 
approximation. To evaluate the role of the outer core in Majoron 
production, we set $\mu_{\nu_e}=\mu_{\nu_\mu}=\mu_{\nu_\tau}=0$.
The density in the outer core drops from $5\times 10^{14}$ g/cm$^3$ to $ 5 
\times 10^{13}$   g/cm$^3$. The temperature in the outer core drops 
abruptly 
\cite{adam} such that 
$T(R=R_{inner})=35$ MeV while $T(R=R_{out})\sim 2$ MeV.

Different models predict different values for parameters; {\it e.g.,} the 
predictions of different classes of models for $T_{inner}$ vary from 10 
MeV 
to 30 MeV \cite{pons}. Moreover the production of Majorons can distort the 
density distributions. Considering these uncertainties, the simplified 
model that we have invoked is justified. With this approach, we will be 
able to examine the prediction of all models for the Majoron luminosity.

\section{ Bounds on coupling constants }
In this section we explore the role of Majorons in the cooling of 
the supernova 
core. In  subsection 4.1, we derive an upper bound on $|g_{ee}|$ 
assuming the produced Majorons leave the core without being trapped.  In 
 subsection 4.2, we derive upper bounds on $g_{\mu \mu}$, $g_{\tau 
\tau}$, $g_{\mu e}$ and $g_{\tau e}$, again assuming Majorons leave the 
core immediately after production.
 In  subsection 4.3, we show that for the couplings lower than the 
bounds we have derived, the four-point interactions are negligible.
 In  subsection 4.4, we derive the limits above which Majorons become 
trapped.
\subsection{Bounds on $|g_{ee}|$}

 As represented in   Table 1, immediately after the  
bounce, $V_{e}$ is positive, but 
eventually $V_{e}$ decreases and 
becomes negative, while $V_{ \mu}$ and $V_{ \tau}$ are  negative 
from the beginning.
As long as $V_{e}>0$, the interactions $\nu_e \rightarrow \bar \nu_e+J$
and $\nu_e \rightarrow \nu_{\mu (\tau)}+J$ are kinematically allowed
but the latter is  suppressed by a factor of $(m/p)^2 \stackrel 
{<}{\sim}10^{-16}$. So we will consider only the interaction
$$ \nu_e \rightarrow \bar \nu_e +J. $$
This interaction depletes the energy of the  core at a rate
\be
{\cal L}_J={|g_{ee}|^2 V_{e} \mu_{\nu_e}^4 \over 12(2 \pi)^3}\times(4/3 
\pi R_{inner}^3).
\ee 
We should note that this interaction not only carries energy away but also 
deleptonizes the core.
\be {d Y_L \over dt}=-2 \Gamma Y_{\nu_e}=-2 {g_{ee}^2 V_{e} \over 
8\pi}Y_{\nu_e}
\ee
where we have used the fact that $n_{\bar\nu_e}\ll n_{\nu_e}$.
We know that the core is in $\beta$-equilibrium. Since the rate of the 
$\beta$-interaction is faster than $\Gamma$  (rate of 
$\beta$-interactions/ $\Gamma \sim 48 \pi G_F^2 \mu_{\nu_e}^3 T^2/ 
g_{ee}^2 
V_e$) and at equilibrium the 
density of electrons is one order of magnitude larger than that of 
neutrinos, we expect that the densities  of the neutrinos are  not 
affected 
by  Majoron production. In other words, the Fermi energy, 
$\mu_{\nu_e}$, and 
$Y_{\nu_e}$ are 
still 
given by Ref. \cite{adam}.
However, deleptonization by Majoron emission can affect $V_{e}$ 
dramatically because different terms in $V_{e}\propto (3 Y_L+Y_\nu-1)/2$
cancel each other $(Y_\nu \ll Y_L\simeq 0.3)$. Therefore in the presence 
of 
Majoron emission, $V_{e}$ vanishes faster.
Let us evaluate the maximum energy that can be carried away by Majorons 
through $\nu_e \rightarrow \bar \nu_e+J$ in the stage when  $V_{e}$ is 
positive. To have an estimation, we can approximate
\be
{d V_{e} \over dt}=-bV_{e}-a,
\label{dvdt}
\ee
where 
$$b=\sqrt{2}{3 \over 8 \pi}G_F {\rho \over M_N} |g_{ee}|^2 Y_\nu$$
and $a$ reflects the 
deleptonization effect without Majoron emission. According to Table 
1,
$a\simeq 2.6$ eV/sec.
If we neglect the variation of $Y_\nu$,  $\rho$
and $a$ with time, we conclude that
$$ V_{e}(t)=(V_{e}(0)+{a \over b})e^{-b t}-{a \over b},$$
so that, after $t_1=(1/b)\times \ln (V_{e}(0)b/a+1)$, $V_{e}$ 
vanishes.
The energy carried away by Majorons up to $t_1$ can be approximated as
\be 
\label{v<0}
E_{V_{e}<0}={g_{ee}^2 \mu_{\nu_e}^4 \over 12 (2\pi)^3}\times 4/3 \pi 
R_{inner}^3\times 
(\frac{V_{e}(0)}{b}-\frac{a}{b^2}\ln {V_{e}(0)b+a \over a}).
\ee
For $g_{ee} \stackrel {>}{\sim} 10^{-7}$, $E_{V_{e}<0}$ converges to 
$4\times 10^{51}$ erg. Increasing 
$g_{ee}$ increases ${\cal L}_J$, but on the other hand, $V_{e}$ 
vanishes in 
a 
shorter period. It
 is 
easy to show that, for any value of $g_{ee}$, 
$$E_{V_{e}<0}<4\times 10^{51} \ \ 
{\rm erg} 
\ll E_b.$$
Therefore the energy loss at this stage does not affect star's 
evolution  and hence we do not obtain any bound.

As  shown in Table 1,  about one second after the core bounce 
$V_{e}$ turns negative. As we discussed earlier, in the presence of 
neutrino 
decay $V_{e}$ changes 
its sign even faster. In a medium with negative $V_{e}$, the decay 
$\nu_e \rightarrow \bar\nu_e+J$ is not kinematically  allowed and 
instead 
$\bar \nu_e \rightarrow \nu_e +J$ can take place. 
However, we know that, in the inner core,
the density of electron antineutrinos is quite low ($\mu_{\bar \nu_e}\sim 
-200$ MeV while $T\sim 10$  MeV)                                          
                                                                    so 
this interaction will not have any role in the cooling of the inner core. 
In 
such a medium, energy  will be 
carried away by process
\be 
\label{taze}
\nu_e+\nu_e \rightarrow J.
\ee
In  previous literature the possibility of this interaction was not 
discussed.
The interaction (\ref{taze}), diminishes the lepton number by two units.
Again we see that $\mu_{\nu_e}$ and $Y_{\nu_e}$ will not be  
considerably 
affected by this process, but that  $V_{e}$ will decrease faster. In 
contrast 
to the previous case, a faster decrease of $V_{e}$ is a positive 
feedback
for the process and leads to the energy depletion.
The energy carried away from the inner core via the process in Eq. 
(\ref{taze}) is now 
\be \label{luminosity}
{\cal L}_J=\frac {7}{12}|g_{ee}|^2 |V_{e}| {\mu_{\nu_e}^4 \over 
(2\pi)^3}\times 
(\frac{4}{3}\pi R_{inner}^3).
\ee

To evaluate a 
conservative upper bound on  $|g_{ee}|$, we set $|V_{e}|$
equal to 0.3 eV, $\mu_{\nu_e}=200$ MeV and $R=10$ km then,
$$ {\cal L}_J=2 |g_{ee}|^2\times 10^{66} \times (\frac {R_{inner}}{10 \ \ 
{\rm 
km}})^3(\frac{V_{e}}{0.3 \ \ eV})(\frac{\mu_{\nu_e}}{200 \ \ {\rm MeV}})^4 
\ \ \ 
\frac{{\rm erg}}{{\rm sec}}
$$ Around one second after the core bounce,  the total 
neutrino luminosity, ${\cal L_\nu}$, is about $5\times 10^{52}$ erg/sec. 
So, the condition ${\cal L}_J<3\times 10^{53}$ erg/sec  yields the 
conservative 
bound, 
 \be
|g_{ee}|< 4 \times 10^{-7}  (\frac {R_{inner}}{10 \ \ {\rm
km}})^{-\frac{3}{2}}(\frac{V_{e}}{0.3 \ \ 
eV})^{-\frac{1}{2}}(\frac{\mu_{\nu_e}}{200 \ \ {\rm MeV}})^{-2}
\label{upper}.
\ee
In Ref. \cite{smirnov}, a bound on $|g_{ee}|$ is obtained by studying the 
energy loss 
via $\bar \nu_e \rightarrow \nu_e +J$ which mainly takes place in the 
outer core, 
$ R_{inner}\simeq 10 \ \ {\rm km} <R<R_{out} \simeq 20 \ \ 
{\rm km}$.
The result is $ {\cal L}(\bar \nu_e \to \nu_e+J)= {\rm few} \times 
10^{64} \ \ |g_{ee}|^2$ erg/sec. So the conservative bound ${\cal 
L}(\bar \nu_e \to \nu_e+J)<3 \times 10^{53}$ erg/sec implies $|g_{ee}|<4 
\times 10^{-6}$. The bound in  Eq. (\ref{upper}) is one order of 
magnitude stronger because the total number of $\nu_e$ in the inner core 
is very high. In Ref. \cite{valle}, a bound is imposed due to the 
 processes
$\nu+\nu \rightarrow J+J$ and $\nu \rightarrow \nu+J$ ($\nu$ denotes both 
neutrino and antineutrino). However the energy carried away is 
overestimated due to an improper treatment of the three-point 
subprocesses. 
We will elaborate on the $\nu+\nu \rightarrow J+J$ process in section 
4.2. 
\subsection{Bounds on $|g_{\mu \alpha}|$ and $|g_{\tau \alpha}|$}
In this subsection we discuss the processes involving $\stackrel 
{(-)}{\nu_\tau}$
and/or $\stackrel {(-)}{\nu_\mu}$. These processes include
$$ (a) \ \ \ \ \ \ \nu_{\mu,\tau}+\nu_{\mu,\tau}\rightarrow J, \ \ \ \ \
\nu_{\mu,\tau}+\nu_e\rightarrow  J $$
and 
$$ (b) \ \ \ \ \ \ \bar\nu_{\mu,\tau}\rightarrow  J+\nu_{e,\mu,\tau}.$$
The process $\bar \nu_{\mu , \tau} \to J +\nu_e$ can take place only in 
the outer core where electron neutrinos are 
not degenerate.
Both  processes (a) and (b) can distort the distribution of matter 
inside 
the star. However, that calculation  is beyond the 
scope of this paper.
But we can argue that it is a good approximation, for the purpose of 
computing upper bounds, to use distributions with vanishing chemical 
potentials for 
$\nu_\tau$ and $\nu_\mu$ 
\cite{adam}. 
 For simplicity we 
rotate $(\nu_\mu,\nu_\tau)$ to a basis such that $g_{\mu \tau}=0$.
Note that since the chemical potential is diagonal and $V_\mu=V_\tau$, it 
will be invariant under this rotation.  
In the new basis, 
we can write, for the inner core
\be \label{evolution} \frac{d n_{\nu_\mu}}{dt}-\frac{d 
n_{\bar\nu_\mu}}{dt}=
2\left[{\rm Rate}(\bar \nu_\mu \rightarrow \nu_\mu +J)-{\rm Rate}(\nu_\mu 
\nu_\mu \rightarrow J)\right]-{\rm Rate}(\nu_\mu \nu_e \rightarrow J),
\ee
where we have neglected $\nu \nu \rightarrow JJ$ interactions.
The sum of the chemical potentials for $\nu_\mu$ and $\bar \nu_\mu$ must 
be 
zero, $\mu\equiv \mu_{\nu_\mu}=-\mu_{\bar \nu_\mu}$,
therefore
\be 
n_{\nu_\mu}=\int \frac{4 \pi} {(2\pi)^3}{p^2 dp \over 
e^{{p-\mu \over T}}+1} \ \ {\rm while} \ \ n_{\bar \nu_\mu}=\int \frac{4 
\pi} 
{(2\pi)^3}{p^2 dp \over
e^{{p+\mu \over T}}+1}. \label{ens}
\ee
We expect that for small values of $|g_{\alpha 
\beta}|$, the chemical potential  remains small. Let us suppose $|\mu/T 
|\ll 1$ to solve 
the 
equation (\ref{evolution}), then we can determine whether this assumption 
is valid or not. For $|\mu/T| \ll 1$, $$n_{\nu_\mu}\simeq {4 \pi T^3 \over 
(2\pi)^3}(1.8+1.64 \mu/T), \ \ \ \ \  n_{\bar \nu_\mu}\simeq (4 \pi T^3 / 
(2\pi)^3)(1.8-1.64 \mu/T)$$ and we can rewrite the right hand side of Eq. 
(\ref{evolution}) as
\be \label{mevolution} {|g_{\mu \mu}|^2 |V_{\mu }| T^3 \over 2 (2\pi)^3}\{ 0.12-3.28 \mu/T-(0.34+0.25 \mu/T)\frac 
{|g_{e 
\mu}|^2}{|g_{\mu \mu}|^2} \frac {|V_{\mu} + V_e|}{|V_\mu|} \frac 
{\mu_{\nu_e}^2}{T^2}\},
\ee
where $\mu_{\nu_e}$ is the  chemical potential of the  electron-neutrinos.
Inside the supernova core, neutrinos and matter are in thermal equilibrium 
and since the energy density of matter is much higher, we expect that the 
rate of thermal change due to these processes is small:
$$|\frac{dT}{T dt}|\sim |{E_{\nu_e}dn_{\nu_\mu} /dt \over E_b/{\rm 
volume}}|\ll 
\frac {dn_{\nu_\mu}}{n_{\nu_\mu} dt}.$$ So, $d 
n_{\nu_\mu}/{dt}-d
n_{\bar\nu_\mu}/{dt}\simeq 2\frac {4 \pi T^3}{(2 \pi)^3}\times 1.64 
{d 
(\mu /T)}/{dt}
$. On the other hand, for this estimation we can neglect the variation in 
$V_{\mu }\simeq \sqrt{2} G_F n_B (Y_e-1)/2$. Also, since the density of 
$\nu_e$ is much higher than that of $\nu_\mu$, we can neglect the 
variation of $\mu_{\nu_e}$.
Therefore, Eq. (\ref{mevolution}) tells us that $\mu/T$ converges to 
$$(0.12-0.34\frac{|g_{e \mu}|^2}{|g_{\mu 
\mu}|^2
}\frac{|V_e+V_\mu|}{|V_\mu|} \frac{\mu_{\nu_e}^2}{T^2})/(3.28+0.25 
\frac{|g_{e \mu}|^2}{|g_{\mu\mu}|^2}\frac{|V_e+V_\mu|}{|V_\mu|} 
\frac{\mu_{\nu_e}^2}{T^2}).$$
Now, it is easy to show that for $(|g_{e \mu}|^2/|g_{\mu \mu}|^2)\times  
(\mu_{\nu_e}^2 /T^2)<37$, $|\mu /T|$ remains small regardless of the 
values of $|g_{e 
\mu}|$ or $|g_{\mu \mu }|$, themselves. For $|g_{e \mu}|>6|g_{\mu 
\mu}|T/\mu_{\nu_e}$, 
$|\mu /T|$ diverges to values larger than 1 and the above analysis is no 
longer correct  (remember that we had assumed $|\mu /T|\ll 1$).
In this case, $\nu_\mu$ will 
disappear after $\sim (\frac {|g_{e \mu}|^2}{100 \pi}V_{\mu 
\mu}\mu_{\nu_e}^2/T^2)^{-1}$ but 
on the other hand, the density of $\bar \nu_\mu$ will increase (the 
chemical potential becomes negative) and this calls for recalculation of 
the density distributions. 
We can make a similar  discussion for $\nu_\tau$. 
Let us  suppose $|g_{e \mu}|<6|g_{\mu \mu}|T/\mu_{\nu_e}$ and $ |g_{e 
\tau}|<6|g_{\tau 
\tau}|T/\mu_{\nu_e}$ and continue from here.

Now let us evaluate the Majoron luminosity using the distributions given 
in \cite{adam}.
Neglecting the Majoron emission from the outer core we find,
\be \label{yek}
{\cal L}(\nu_\alpha+\nu_\beta \rightarrow J)\simeq \frac{2}{3} 
\frac{R_{inner}^3}{(2 \pi)^2} T_{inner}^4 
|g_{\alpha \beta}|^2(|V_{ \alpha}+V_{ \beta}|)
\ee
and
\be\label{do}
{\cal L}(\bar \nu_\alpha\rightarrow \nu_\beta  J)\simeq 
\frac{1.3}{3}\frac{R_{inner}^3}{(2 \pi)^2} T_{inner}^4 
|g_{\alpha \beta}|^2(|V_{ \alpha}+V_{ \beta}|),
\ee
where by $\alpha$ and $\beta$ we denote $\mu$ or $\tau$.
On the other hand,
\be 
{\cal L}(\nu_{\alpha}+\nu_e \rightarrow J)
\simeq {R_{inner}^3 \over 3 (2 
\pi)^2}
|g_{\alpha e}|^2(|V_{ \alpha}+V_{e}|)(0.2 
\mu_{\nu_e}^3 T_{inner}).
\label{se}
\ee
Note that even if $g_{\alpha e}$ is large, the process ( $\bar 
\nu_{\alpha} 
\rightarrow \nu_e +J$), 
 in the inner core, is suppressed by a factor of 
$exp((T_{inner}-\mu_{\nu_e})/T_{inner})$ 
because  
inside the star, $\nu_e$ is degenerate.

Then, the requirement ${\cal L}<3 \times 10^{53}$ erg/sec implies
\be 
\label{bonmu}
\sqrt{
\sum_{\alpha , \beta \in \mu,\tau} |g_{\alpha 
\beta}|^2}\stackrel 
{<}{\sim}8 \times 10^{-7} ({10\ \ {\rm eV} \over |V_{ 
\mu}|})^{\frac{1}{2}} 
({20 \ \ {\rm MeV}\over T_{inner}})^2 ({10 \ \ {\rm km}\over 
R_{inner}})^{\frac 
{3}{2}} 
\ee
and 
\be 
\label{bone}
\sqrt{|g_{\mu e}|^2+|g_{\tau e}|^2}<5\times 10^{-7}({10 \ \ {\rm km} \over 
R_{inner}})^{\frac{3}{2}}({200 \ \ {\rm MeV} \over 
\mu_{\nu_e}})^{\frac{3}{2}} 
({20 \ \ 
{\rm 
MeV} \over T_{inner}})^{\frac{1}{2}} ({10 \ \ {\rm eV} \over V_{ 
\mu}})^{\frac 
{1}{2}}.
\ee 
We emphasize again that the above results are valid only  
assuming
that $|g_{e \mu}\mu_{\nu_e}/g_{\mu \mu}T|^2<37$ and $|g_{e 
\tau}\mu_{\nu_e}/g_{\tau \tau}T|^2<37$. Otherwise the 
$\nu_{\mu(\tau)}$ annihilation will stall because $\nu_{\mu(\tau)}$ is 
depleted. Meanwhile, the energy carried away due to 
$\nu_{\mu(\tau)}$-annihilation is of the order of 
$${\cal L}(\nu_e +\nu_{\mu(\tau)} \rightarrow J)(\frac{|g_{e 
\alpha}|^2}{100 
\pi} 
V_{\mu } \frac{\mu_{\nu_e}^2}{T^2})^{-1}\sim 10^{49} \ \ {\rm erg} \ll 
E_b\sim 
10^{53}  \ \ {\rm 
erg}$$
So,  ${\cal L}(\nu_e+\nu_{\mu(\tau)} \rightarrow J)$ does not impose any 
bound 
on $|g_{e \mu (\tau)}|$.
On the other hand, since the density of $\bar \nu_{\mu(\tau)}$ grows, the 
process ${\cal L}(\bar \nu_{\mu(\tau)}\rightarrow \nu_{\mu(\tau)}J)$ will 
even become intensified and we expect that still Eq. (\ref{bonmu}) will be 
a conservative  bound. 
For $(|g_{e \mu}|^2/|g_{\mu \mu}|^2)\times
(\mu_{\nu_e}^2 /T^2)>37$, the upper bound on $|g_{e \mu}|$ is imposed by 
$\bar \nu_\mu$-decay in the outer core. Using the distributions in  
Ref. \cite{adam}, we can show that \be \label{cha}{\cal L}(\bar 
\nu_{\mu (\tau)} \to 
J+\nu_e)={\rm 
few} \times |g_{e \mu (\tau)}|^2 \times 10^{64}\ \ {\rm 
erg/sec} \ee which implies \be 
\label{outcore} |g_{e 
\mu}|,|g_{e \tau}|<{\rm few} \times 10^{-6}.\ee In  Figs. (1) and (2), all 
these 
bounds are schematically
depicted for $T_{inner}=10$ MeV and $T_{inner}=20$ MeV, respectively. The 
shadowed area represents the range of parameters for which ${\cal 
L}_J<3 \times 10^{53}$ erg/sec. As  shown in Fig. (4), for $T=20$ 
MeV, the process 
$\nu_e+\nu_\mu\to J$ does not impose any bound on $|g_{e \mu}|$ because, 
for any 
value of $|g_{e \mu}|$ smaller than $\sqrt{37} |g_{\mu \mu}|T/\mu_{\nu_e}$
(where $|g_{\mu \mu}|$ is below its upper bound) it cannot  give rise 
to a
Majoron luminosity larger than the allowed value.

\subsection{Four-point interactions}
  In this subsection we discuss the processes 
$\nu+ \nu \rightarrow J+J$ and $\bar 
\nu+\bar \nu 
\rightarrow J+J$. As discussed in sections 2.2.3 and 2.2.4, we consider 
only the intrinsically connected contributions, the effects of 
three-particle sub-processes subtracted.
Using the distributions in Ref. \cite{adam} and the formulae we have found 
in 
 subsection 2.2.3,  we obtain
\be
{\cal L}(\nu_e+\bar \nu_{\mu (\tau )}\to J+J)\sim 
\frac{\mu_{\nu_e}^3 
T_{inner}^2}{(2 
\pi)^4}(\frac 
{4 \pi} {3} R_{inner}^3)| \sum_{\alpha} g_{e \alpha} g^*_{\mu ( \tau) 
\alpha} |^2
\label{one}
\ee
and
\be \label{two}
{\cal L}(\nu_{\mu (\tau )}+\bar \nu_{\mu (\tau )}\to J+J)\sim
  \frac{T_{inner}^5}{(2\pi)^4}(\frac{4 \pi} {3} R_{inner}^3)| 
\sum_{\alpha} g_{\mu (\tau ) \alpha} g_{\mu (\tau ) \alpha}^* |^2. 
\ee
In the above equations, $\alpha$ runs over $\{ e, \mu , \tau \}$.
Using the distributions in Ref. \cite{adam} and the formulae we found in
 section 2.2.4,  we obtain
\be \label{three}
{\cal L}(\nu_e+\nu_e \rightarrow 
J+J)=\frac {1}{(2\pi)^6} \left( 
a\frac {m^2} {V_{\mu}^2/10}|g_{e \mu}^2+g_{e 
\tau}^2|^2+b|g_{ee}|^4 \right) (\frac{4\pi}{3}R_{inner}^3)\mu_{\nu_e}^5, 
\ee
\be 
\label{four}
{\cal L}(\nu_e+\nu_{\mu (\tau )} \rightarrow 
J+J)=
\ee 
$$(\frac{4\pi}{3}R^3){\mu_{\nu_e}^3T_{inner}^2 \over (2 \pi)^6}
\left( a' | g_{e \mu} g_{\mu \mu (\tau)}+g_{e \tau} g_{\tau \mu(\tau)}|^2 
\frac {m^2}{V_{\mu}^2/10}+b'|g_{ee}g_{e \mu (\tau)}|^2 \right)
$$
and
\be
{\cal L}(\nu_\alpha+\nu_\beta \rightarrow J+J) \sim
{\cal 
L}(\bar \nu_\alpha+\bar \nu_\beta \rightarrow J+J)= \ee $$
(\frac{4\pi}{3}R_{inner}^3){T_{inner}^5 \over (2 \pi)^6}
\left( a'' |g_{\mu  \mu (\tau)}^2+g_{\mu \mu (\tau)}^2|^2+ b'' |g_{e 
\mu (\tau)}|^2 \right),
$$
where $m$ is the neutrino mass for quasi-degenerate mass 
schemes.
If $|g_{\alpha \beta}|$ is smaller than the ``upper" bounds in  Eqs. 
(\ref{upper}, \ref{bonmu}, \ref{outcore}) the above luminosities are 
negligible. 
These luminosities become non-negligible  only if the couplings 
are  larger than $10^{-5}$ so they do not change the ``upper" bounds.
Eqs. (\ref{one}-\ref{four}) depend on  combinations of 
couplings so in general it is rather difficult to 
compare them 
with the 
Majoron luminosity due to three-point processes ({\it i. 
e.,} ${\cal L}(\bar \nu_\alpha \to J \nu_\beta)$, ${\cal 
L}( \nu_\alpha\nu_\beta\to J)$).
Let us suppose that all elements are zero except for a 
particular $g_{\alpha \beta}$. Then, Eqs. 
(\ref{yek},\ref{do},\ref{two},\ref{four}) imply that for 
$|g_{\mu \mu}|<5\times 10^{-3}$ the three-point processes 
are dominant. If all couplings, but $|g_{ee}|$ are zero Eqs. 
(\ref{luminosity}) and (\ref{three}) show that as long as 
$|g_{ee}|<10^{-3}$, $\nu_e \nu_e \to J$ is dominant.
Comparing Eqs. (\ref{one}) and (\ref{three}) with Eq. 
(\ref{cha}) reveals that as long as $|g_{e \mu}|<10^{-4}$ 
the process $\bar \nu_\mu \to \nu_e J$ dominates.  
We note that, for coupling constants of the order of the ``lower" bound 
(for which 
the produced Majorons are trapped), the four-point processes can play a 
significant role.
\subsection{Majoron decay and scattering}
So far we have assumed that Majorons leave the star without undergoing any 
interaction or decay. Now we discuss the validity of this assumption.
First, let us discuss the possibility of decay. (Note that although 
Majorons are massless particles, in a medium such as a supernova, in 
principle,  they can 
decay.) For $\alpha , \beta \in \{ \mu , \tau \}$,
\be \label{jdecays}
\Gamma(J(q)\rightarrow \nu_\alpha+\nu_\beta)={|g_{\alpha \beta}|^2(|V_{ 
\alpha}+V_{\beta }|) \over 8 \pi}(0.8-0.27),
\ee 
where 0.8 and 0.27 correspond to $q/T=10$ and $q/T=0.1$, respectively. 
So, the Majorons decay before leaving the core  ($\Gamma >  1/R$),
only if
\be
\label{decay}
|g_{\alpha \beta}|\stackrel {>} {\sim} 10^{-5}.
\ee

Because of degeneracy of the inner core, only the energetic Majorons 
$(|E_J-\mu_{\nu_e}|/ T\stackrel {>}{\sim} 1)$ can decay into electron 
neutrinos ({\it see} Eq. 
(\ref{Jdecay})). It can be shown that 
 $$\Gamma[J(q>2\mu_{\nu_e})\rightarrow \nu_e+\nu_e]\sim {T|g_{ee}|^2 |V_e| 
\over 
4 \pi \mu_{\nu_e}}$$ and  $$\Gamma[J(q>\mu_{\nu_e})\rightarrow 
\nu_e+\nu_\alpha]\sim {|g_{ee}|^2 |V_e+V_\mu| \over 8 \pi }.$$ 
If
\be \label{gemu}
|g_{e \mu}| >7 \times 10^{-6} \ \ \ \ {\rm and/or} \ \ \ \ |g_{ee}|>5 
\times 10^{-5}.
\ee
Majorons that are produced in the center will decay before leaving the 
core. Note that, even beyond the neutrinosphere  as long as 
$|V|$ is large enough, 
 Majoron decay can take 
place. 
Now let us examine the interaction effect. For low values of coupling 
constants, the dominant interactions are 
$(\nu + J \rightarrow \bar \nu)$ with the mean free path
\be l^{-1}(\nu_e+J\rightarrow \bar \nu_e)=\frac{|g_{e 
e}|^2}{4 
\pi} \frac{\mu_{\nu_e}}{q}|V_{e}|\label{makhraj},
\ee
\be
l^{-1}(\nu_e  + J \rightarrow \bar \nu_\beta)= \frac{|g_{e 
\beta}|^2}{8 \pi} \frac{|V_{e}+V_{ \beta}|}{q} 
(\mu_{\nu_e}-T \ln({e^{q/T}+1 \over e^{q/T}})), \label{jinter}\ee

 \be \label{jinteracts} l^{-1}(\nu_\beta +J\rightarrow \bar \nu_\alpha)=
\frac{|g_{\beta\alpha}|^2}{8 \pi} \frac{1} {q}(|V_{ 
\beta}+V_{\alpha }|){T e^{q/T}(q/T+\ln 2 -\ln(e^{q/T}+1) \over 
e^{q/T}-1}
\ee

and

\be
l^{-1}(\nu_\beta +J\rightarrow \bar \nu_e)=\frac{|g_{\beta e}|^2}{8 
\pi} \frac{0.7 T}{q}(|V_{ \beta}+V_{e}|),
\ee
where $q$ is the energy of $J$ and $\alpha$ and $\beta$ are either 
$\mu$ or $\tau$.
The requirement $l^{-1}>R^{-1}$ implies that 
$$|g_{ee}|\stackrel {>}{\sim}6\times 10^{-6} (\frac {q}{10 \ \ {\rm 
MeV}})^{\frac{1}{2}}(\frac{200 \ \ {\rm 
MeV}}{\mu_{\nu_e}})^{\frac{1}{2}}(\frac 
{0.3 \ \ {\rm eV}}{|V_{e}|})^{\frac{1}{2}},$$
$$|g_{e \mu}|,|g_{e \tau}|\stackrel {>}{\sim}2\times 10^{-6} (\frac {q}{10 
\ \ {\rm
MeV}})^{\frac{1}{2}}(\frac{200 \ \ {\rm 
MeV}}{\mu_{\nu_e}})^{\frac{1}{2}}(\frac
{10 \ \ {\rm eV}}{|V_{ \mu}|})^{\frac{1}{2}}$$
and 
$$|g_{\mu \mu}|,|g_{\tau \tau}|, \sqrt{2}|g_{\tau \mu}|\stackrel 
{>}{\sim}4 \times 10^{-6}.$$
In the last case, the bound is derived for $q=10$ MeV, 
$T=10$ MeV and $|V_{ \mu}|=10$ eV.
Note that these bounds are derived for the 
parameters inside the inner core. In the outer core $\mu_{\nu_e}$ is much 
smaller and therefore the mean free path is much larger, {\it i.e.,} in 
the 
outer core neutrinos can escape more easily.
 Apparently if the coupling constants are smaller than   the bounds in Eqs.
(\ref{upper},\ref{bonmu},\ref{bone}),  Majorons will leave the star 
core before 
undergoing any interaction. 
For $6 |g_{\mu \mu}| (T/\mu_{\nu_e})<|g_{e \mu}|\sim 2 \times 
10^{-6}$, although the Majorons produced in the outer core escape 
immediately (recalling  that the bound in Eq. (\ref{outcore}) is 
extracted by studying the process $\bar \nu_\mu \to \nu_e+J$ in the outer 
core), the interaction of Majoron particles with $\nu_e$ in the inner 
core is not negligible.\newline For larger values of coupling, Majorons 
may become  trapped or decay before leaving the star and the energy 
transfer by Majoron emission will become harder, but this does not mean 
that Majoron production does not affect the supernova evolution. To 
calculate   the exact effect and to extract lower bounds on coupling 
constants, one needs to  revisit the matter distribution and its time 
evolution including the effect of energy transfer by Majorons. 
That is beyond the scope of this paper. Here, we have   discussed only the 
dominant interaction modes for larger values of the coupling constants.
We recall that for $|g|\stackrel {>}{\sim} 5\times 10^{-4}$ the effective 
mass of the Majoron becomes non-negligible.

\section{Conclusions and discussions}
We have explored the energy loss from the inner core due to emission of 
Majorons. We have found that at  early instants after the shock bounce 
($t\stackrel {<}{\sim}1$ 
sec) when $V_e$ is positive, although the decay $\nu_e \rightarrow \bar 
\nu_e +J$ takes place, the period is too short to have 
significant energy transfer and therefore the energy loss due to  $\nu_e 
\rightarrow 
\bar \nu_e+J$ does not imply any bound on $|g_{ee}|$. In the next period 
$(t>1$ sec) when $V_e<0$, neutrino decay is no longer kinematically 
allowed 
and instead the two processes $\bar \nu_e \rightarrow \nu_e+J$ and 
$\nu_e+\nu_e 
\to J$ can take place.
Since  the density of $\nu_e$ is much higher in the inner core,
the process $\nu_e+\nu_e \to J$ implies a stronger bound. We have 
found that $|g_{ee}|<10^{-7}$ ({\it see} 
Eq. (\ref{upper})) if the emitted Majoron leaves the core immediately 
after production. 
We have found that for $|g|<10^{-7}$ the effect of four-point 
processes 
($\nu_e+\nu_e \to J+J$ and $\bar \nu_e+\nu_e \to J+J$) 
is negligible. We believe that previous treatments of the reaction 
({\it e.g.,} \cite{valle}) have not correctly subtracted the 
3-particle subprocesses.

We also have studied the bounds on coupling of Majoron to muon (tau) 
neutrino.
In the basis in which $g_{\mu \tau}=0$ (since $\nu_\mu$ and $\nu_\tau$ 
are equivalent for supernova processes, we can always rotate ($\nu_{\tau}, 
\nu_{\mu})$ to a new basis ($ \nu_{\tau}',\nu_{\mu}')$ for which 
$g_{ \nu_{\tau}' \nu_{\mu}'}=0$) we have found the following results.  
For $|g_{e \mu}|^2/|g_{\mu 
\mu}|^2\times \mu_{\nu_e}^2/ T^2<37$, 
the processes $\bar \nu_\mu \to \nu_\mu+J$ and $\nu_\mu+\nu_\mu \to J$ 
imply
$|g_{\mu \mu}|<8 \times 10^{-7}$ ({\it see } Eq. (\ref{bonmu})) while 
$\nu_e 
+\nu_\mu \to J$ gives $|g_{e \mu}|<5 \times 10^{-7}$ ({\it see } Eq. 
(\ref{bone})),
 providing the 
emitted Majoron leave the core without being trapped or 
undergoing decay.
\newline
 For $|g_{e \mu}|^2/|g_{\mu \mu}|^2\times (\mu_{\nu_e}^2)/ T^2>37$, we 
have shown 
that the process $\nu_e+\nu_\mu \to J $ eats up $\nu_\mu$ within a short  
period (leading to a negative  chemical potential for  $\nu_\mu$)
such that the bound from $\nu_e+\nu_\mu \to J$  no longer applies. 
However, in this case, the density of $\bar \nu_\mu$ increases ($\mu_{\bar 
\nu_\mu}=-\mu_{\nu_\mu}$ becomes positive) and the bound on $|g_{\mu 
\mu}|$ 
(Eq. (\ref{bonmu})) still applies (actually it will be a conservative 
one).
For $|g_{e \mu}|>|g_{\mu \mu}| \sqrt{37}T/ \mu_{\nu_e}$,
the $\bar \nu_\mu$-decay in the 
outer core (where $\mu_{\nu_e}/T \stackrel {<}{\sim} 1$) imposes the 
strongest bound on $|g_{e \mu}|$ which is  $ |g_{e \mu}|<{\rm few}\times 
10^{-6}$. 
 These upper bounds are schematically summarized in 
Figs. (1) and (2).
Note that the bounds on $|g_{\tau \tau}|$ and $|g_{e \tau}|$ are exactly 
the same as those on $|g_{\mu \mu}|$ and $|g_{e \mu}|$, 
respectively.
The bounds are parameterized in terms of supernova 
parameters ($T$, chemical potentials, $V$ and the radius of 
the  core) so it is possible to apply
 the predictions of 
any supernova model. \newline
All these upper bounds come from three-particle processes shown in 
diagrams (a,b) of Fig. (2). In these 
processes,
all the involved particles are on-shell. Therefore  the aforementioned 
bounds can be translated into  bounds on the corresponding element of 
the matrix $h$  in the derivative form of interaction (see Eq. 
(\ref{binteraction})), using the relation given in Eq. (\ref{replace}).

We also studied Majoron decay and the interactions that can trap Majorons.
We have found that the processes $\nu_e +J \to \bar \nu_e$, $\nu_e +J \to 
\bar \nu_{\mu,\tau}$ and $ \nu_{\mu,\tau}+J \rightarrow \bar 
\nu_{\mu,\tau}$ may have significant effect ($l^{-1}>R_{core}^{-1}$), only  
if 
$|g_{ee}|> 6 \times 10^{-6} (q/ {\rm 10 \ \ MeV})^{1/2}$, 
$|g_{e 
\mu(\tau)}|>2\times 10^{-6} (q/ {\rm 10 \ \ MeV})^{1/2}$ 
and $|g_{\mu 
(\tau)\mu (\tau)}|> 4 \times 10^{-6}$, respectively. 
If the couplings of Majorons to neutrinos are larger than these limits, 
 the Majorons cannot leave the core immediately. However, the 
processes involving  the Majoron still affect the evolution of supernova, 
transferring 
energy from the inner core  and distorting  the density 
distribution of the particles. If the couplings of Majoron are larger than 
some  
lower bounds, the only Majoron particles that can leave the core and cool 
down it are those produced in (or diffused into)  a shell close to the 
neutrino-sphere where the density decreases rapidly  with 
increasing radius. In this 
region the density is too low to give rise to a significant Majoron flux 
({\it i.e.,} ${\cal L}_J \ll{\cal L}_\nu$). We emphasize that to derive 
the lower bounds, it is not sufficient  to consider the coupling constants 
collectively. For example, if $|g_{e \mu}|>5 \times 10^{-6}$, the Majorons 
produced via $\nu_e+\nu_e \to J$ can annihilate with another $\nu_e$ into 
$\bar \nu_\mu$ before escaping the core.
\newline  
 To derive the lower 
bounds, 
one must recalculate
the density and temperature profiles of matter, neutrinos and Majoron 
particles which, in general, are different from those calculated so far 
without including Majoron processes.
Here, we have  discussed and evaluated only the four-point interactions 
which  for large values of coupling constants may have significant effect.
\section*{ Acknowledgment}
I would like to thank J. Beacom, N. F. Bell, A. Burrows, H. Quinn and A. 
Yu. Smirnov 
for useful comments. I am also especially grateful to Michael Peskin for 
fruitful discussions and encouragement.

\newpage
\renewcommand{\baselinestretch}{0.5}
\begin{figure}
\begin{center}
\vskip -3.2cm 
\hskip -7.0cm
\parbox[c]{3.5in}
{\mbox{
\qquad\epsfig
{file=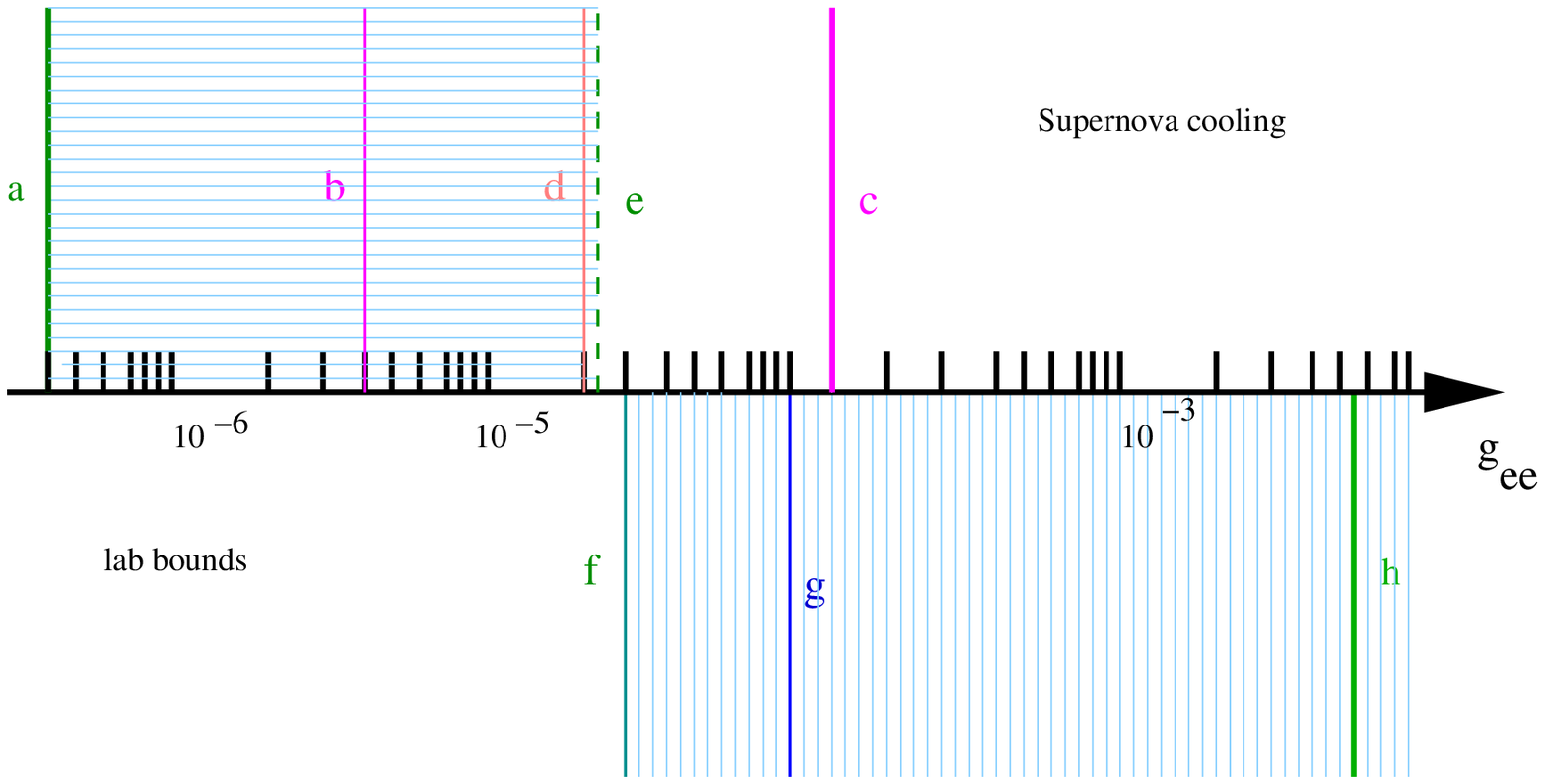,width=1.6\linewidth,height=1\linewidth}}}   
\caption{Bounds on $g_{ee}$ from supernova cooling  (upper 
lines) and lab observations (lower lines).  
Line (a) shows the ``upper" bound on 
$g_{ee}$ 
from the
$\nu_e \nu_e\to J$ process in the supernova core (see Eq. 
(\ref{upper})) while
line (b) represents the ``upper" bound  from 
$\bar \nu_e \to \nu_e J$ \cite{smirnov}. 
 Line (c) shows the ``lower" bound 
 which is derived without considering the 
effect of the four-point processes \cite{smirnov}. 
Line (c) gives the ``lower" bound
according to 
\cite{valle}, we have argued that this is an overestimation. Thus we 
expect the true ``lower" bound (e) 
to be between lines (c) and (d). The range of parameters between the 
``upper" and 
``lower" bounds (the horizontally shaded area) is excluded by supernova 
considerations. 
Line (f) represents the upper bound from double beta decay 
\cite{betabeta} and the whole region to its  right (the vertically 
shaded area) 
is 
excluded.
 Lines  (g) and (h) represent the upper bounds 
derived from   solar 
neutrinos \cite{beacom} and Kaon decay \cite{pika}, 
respectively.  Note that the 
bound (f) from solar data applies to $g_{21}$ rather than  
$g_{ee}$; we have included this line  to compare the 
orders of magnitudes of different bounds. We have not resolved whether the 
``lower" bound (e) lies above or below (f); in the latter case, there is a 
small allowed region between the two bounds. 
} \end{center} \end{figure}
\newpage \begin{figure}
\begin{center}
\vskip -3.2cm 
\hskip -7.0cm
\parbox[c]{3.5in}
{\mbox{
\qquad\epsfig
{file=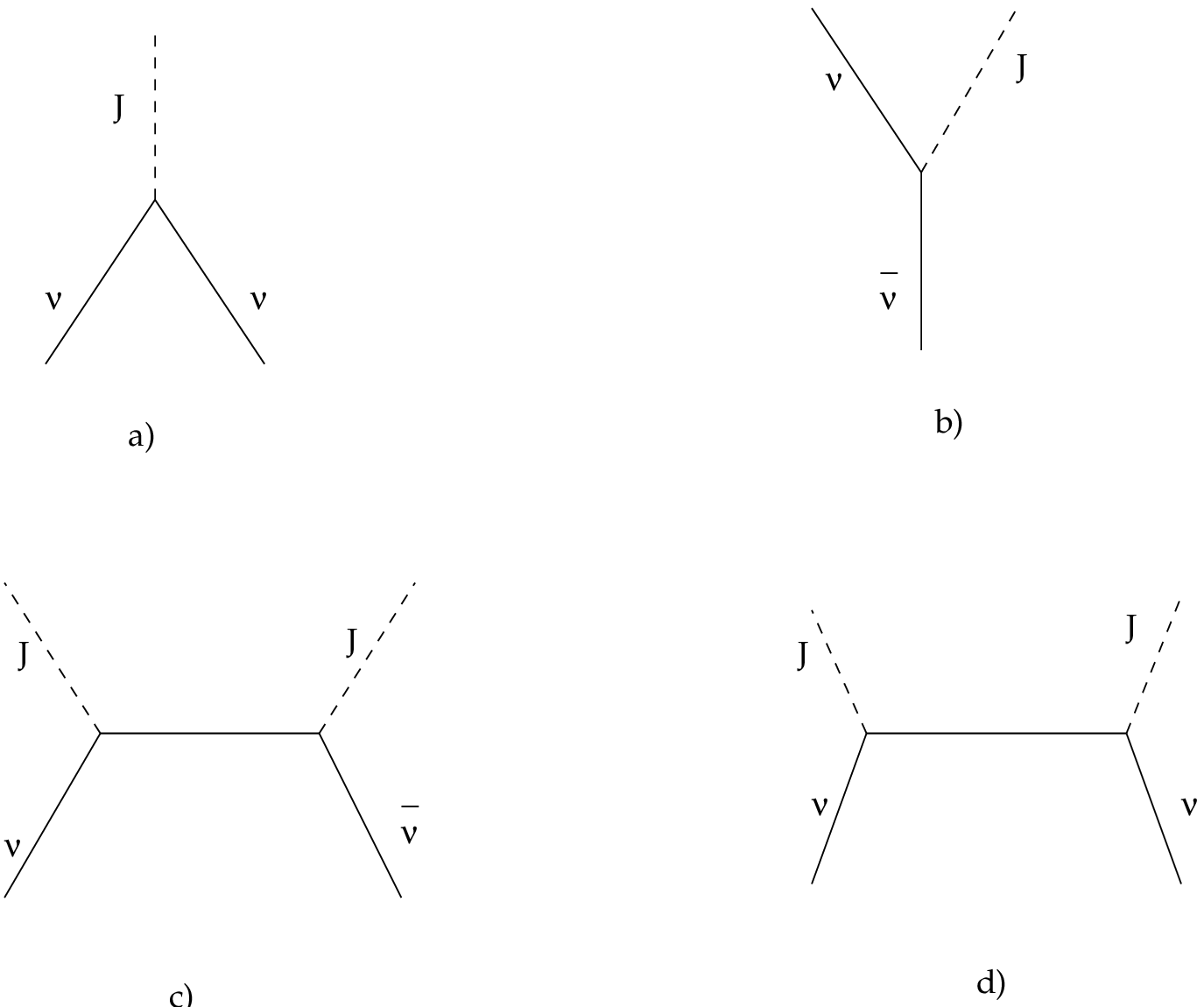,width=1.6\linewidth,height=1.3\linewidth}}}   
\vskip 1 cm
\caption{
Diagrams (a) and (b) are the dominant three-point processes  and are 
possible only for $V<0$.
Diagrams (c) and (d) are the subdominant diagrams and  can take place 
for any value of $V$.}
\end{center}
\vskip -0.8 cm
\end{figure}
\newpage \begin{figure}
\begin{center}
\vskip -3.2cm 
\hskip -7.0cm
\parbox[c]{3.5in}
{\mbox{
\qquad\epsfig
{file=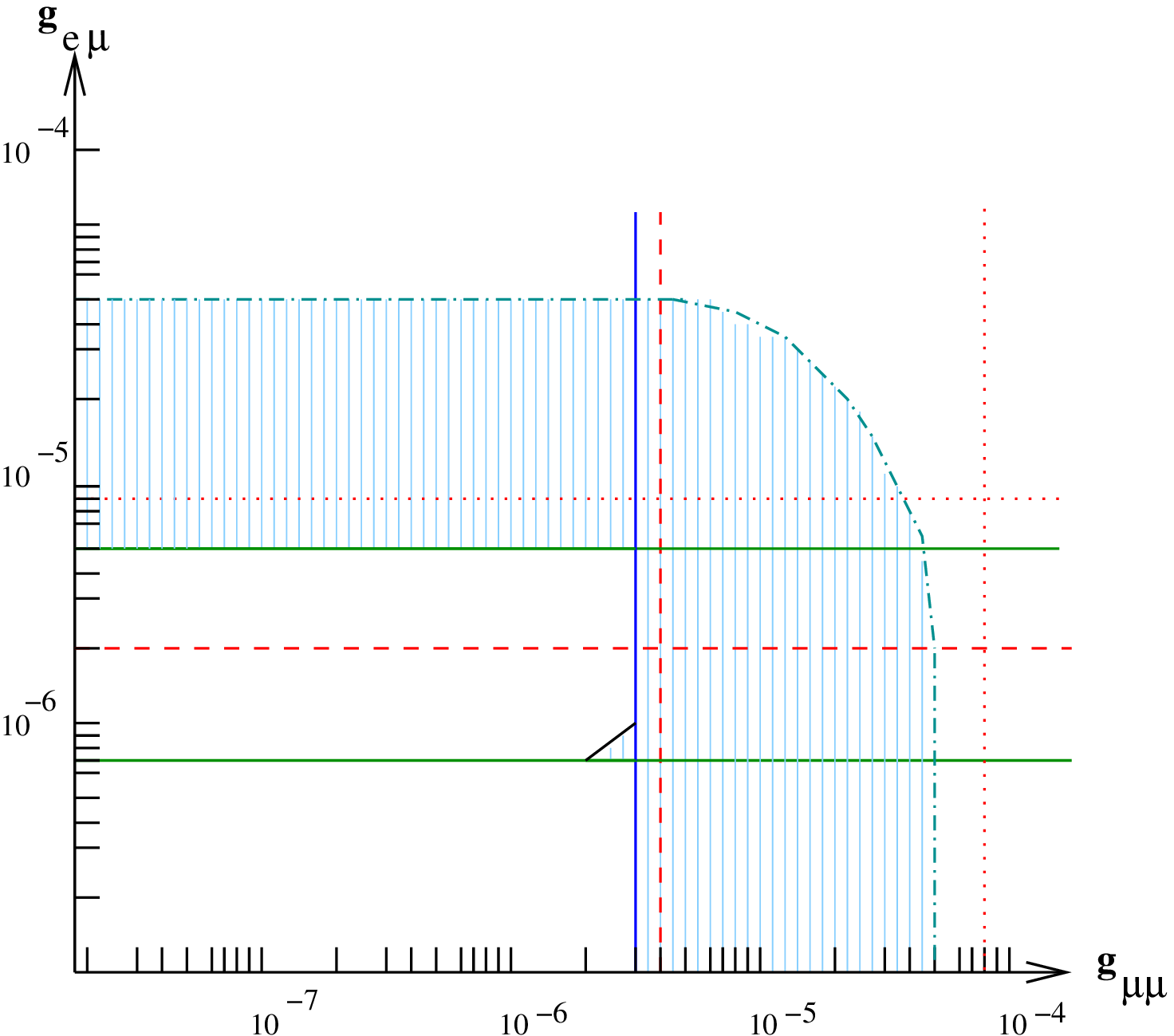,width=1.6\linewidth,height=1.5\linewidth}}}   
\vskip 1 cm
\caption{The bounds on coupling constants for $T=10$ MeV and
$\mu_{\nu_e}=200$ MeV. The shaded area is excluded by energy loss considerations. 
The horizontal and vertical lines at $5\times 
10^{-6}$ and $3\times 10^{-6}$ represent the upper bounds obtained in Eq. 
(\ref{outcore}) and Eq. (\ref{bonmu}), respectively. 
The dashed lines show the limits above which Majorons with
energy $\sim 10$ MeV scatter before leaving the core. The dotted
lines represent the same limits for Majorons with energy $\sim 200$ 
MeV (see Eqs. (\ref{jinter},\ref{jinteracts})). The dot-dashed line 
schematically
represents the ``lower" bound. We have not calculated
the exact numerical value of the lower bound, but this is an estimate  
for $g_{ee}=0$. 
Note that the energies of Majorons produced via $\nu_\mu \nu_\mu \to J$ 
and 
$\bar \nu_\mu \to J \nu_\mu$ are of the order of 10 MeV; that is why the 
``lower" bound can be to    the left  of the vertical dotted line.  
The values inside the notch (above the horizontal line at $7\times 
10^{-7}$) are excluded due to $\nu_\mu \nu_e \to J$ (see Eq. (\ref{bone}) 
and its discussion).
}
\end{center}
\vskip -0.8 cm
\end{figure}
\newpage
\begin{figure}
\nonumber
\begin{center}
\vskip -3.2cm
\hskip -7.0cm   
\parbox[c]{3.5in}
{\mbox{
\qquad\epsfig
{file=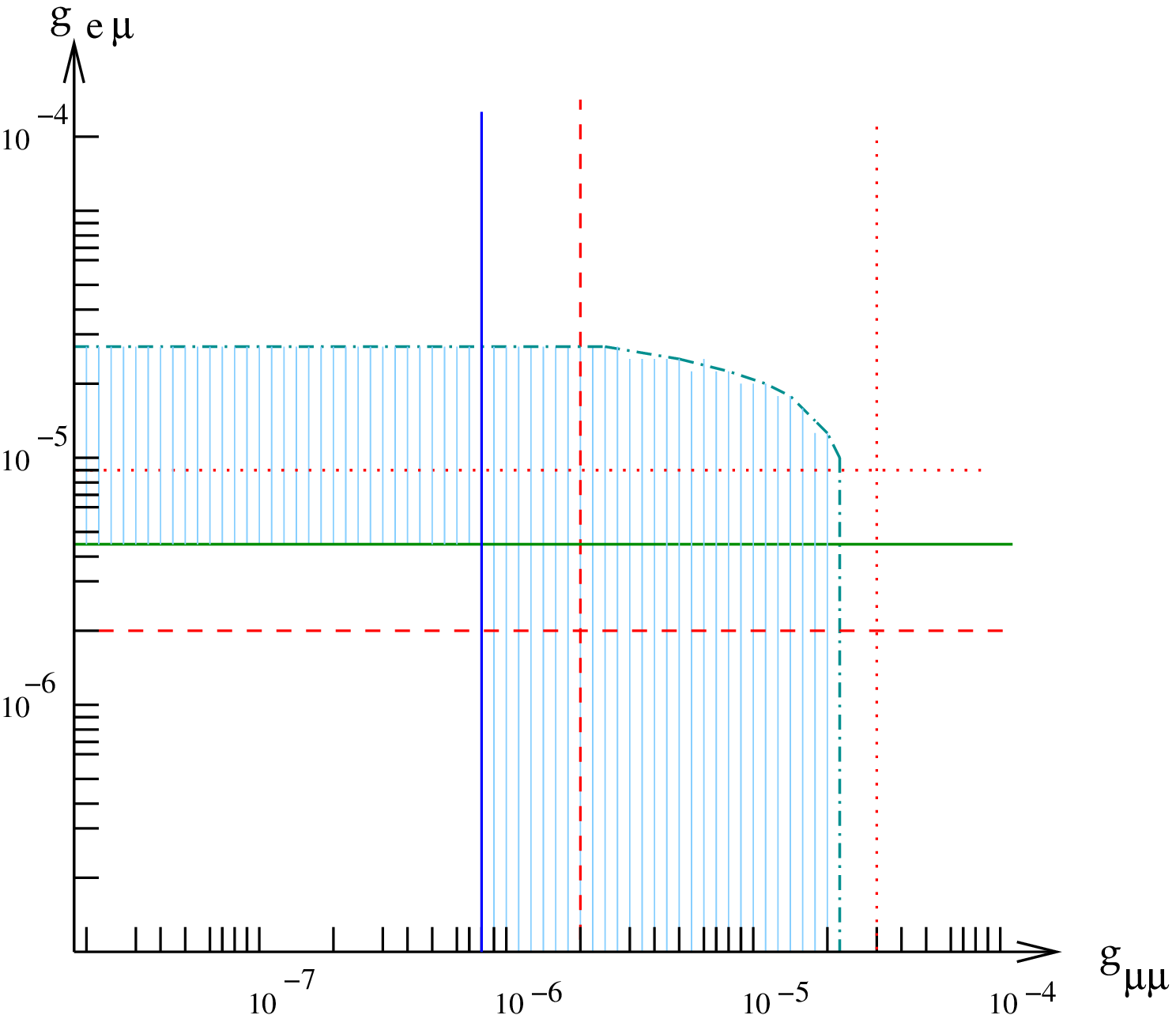,width=1.6\linewidth,height=1.5\linewidth}}}
\vskip 1 cm
\caption{The bounds on coupling constants for $T=20$ MeV and 
$\mu_{\nu_e}=200$ MeV. 
The shaded area is excluded by energy loss considerations.
The horizontal and vertical lines at $4.5\times
10^{-6}$ and $8\times 10^{-7}$ represent the upper bounds obtained in Eq.
(\ref{outcore}) and Eq. (\ref{bonmu}), respectively.
The dashed lines show the limits above which Majorons with
energy $\sim 10$ MeV scatter before leaving the core. The dotted
lines represent the same limits for Majorons with energy $\sim 200$
MeV (see Eqs. (\ref{jinter},\ref{jinteracts})). The dot-dashed line
schematically
represents the ``lower" bound. We have not calculated
the exact numerical value of the lower bound, but this is an estimate 
for $g_{ee}=0$.
Note that the energies of Majorons produced via $\nu_\mu \nu_\mu \to J$ 
and
$\bar \nu_\mu \to J \nu_\mu$ are of the order of 10 MeV; that is why the
``lower" bound can be to   the left  of the vertical dotted line.
}
\end{center}

\vskip -0.8 cm
\end{figure}


\end{document}